\documentclass[10pt, a4paper]{article} 

\usepackage[english]{babel} 
\usepackage{booktabs} 
\usepackage{comment} 
\usepackage[utf8]{inputenc} 
\usepackage[T1]{fontenc} 
\usepackage{amsmath,amsfonts,amsthm,amssymb}
\usepackage{tikz}
\usepackage{tikz-cd}
\usetikzlibrary{positioning,arrows}
\usetikzlibrary{decorations.pathreplacing}
\usepackage{tikzsymbols}
\usepackage{blindtext}
\usepackage{textgreek}
\usepackage{adjustbox}
\usepackage{xcolor}
\usepackage{etoolbox}
\usepackage{url}
\usepackage{gensymb}

\usepackage{caption}
\usepackage{subcaption}

\usepackage{geometry}

\usepackage{setspace}
\usepackage[T1]{fontenc}
\usepackage[utf8]{inputenc}

\usepackage{fancyhdr}
\usepackage{natbib}

\title{\textbf{The Green Peace Dividend: the Effects of Militarization on Emissions and the Green Transition}}
\author{\scalebox{1.15}{Balázs Markó\thanks{I would like to thank Valentina Bosetti, Basile Grassi, Massimo Morelli, Dmitriy Sergeyev, Marta Prato, Luigi Iovino, Carlo Favero, Gianmarco Ottaviano, and Darius Corbier for their helpful comments. I would also like to thank the Numbers and Stories 2 conference and the II Political Economy Winter Workshop participants for their comments. Balázs Markó: balazs.marko@phd.unibocconi.it}}\\ \scalebox{1.1}{Bocconi University}}
\date{\scalebox{1.15}{January 2025}}

\begin{document}

\maketitle

\begin{abstract}
\noindent
    This paper argues that military buildups lead to a significant rise in greenhouse gas emissions and can disrupt the green transition. Identifying military spending shocks, I use local projections to show that a percentage point rise in the military spending share leads to a 0.9-2\% rise in total emissions, as well as a 1\% rise in emission intensity, depending on the economy's overall emission intensity. Using a dynamic production network model calibrated for the US, I find that a permanent shock of the same size would increase total emissions by between 0.36\% and 1.81\%, and emission intensity by between 0.22\% and 1.5\%. The empirical analysis indicates that green patenting is reduced by 10-25\% following such a shock, and the model suggests that investment in renewables could be crowded out by defence spending under certain circumstances, hindering the energy transition. These effects can significantly raise climate damages and temperatures. Depending on the social cost of carbon and the composition of a military spending shock, I estimate that doubling the military spending share in the US in 2017 (equal to 3.3\% of GDP) would have led to climate damages equivalent to between 0.07-2.6\% of GDP per year.
\end{abstract}

\setcounter{page}{1}

\section{Introduction}

After several decades of relative calm and peace in Europe, a flare-up of wars and a souring of international relations between the major powers of the world prompted economists and policymakers to redirect their attention toward understanding the effects of military activity.

 The risk of escalation in specific conflicts, such as the Ukraine war, or the Israel-Hamas war, raises the likelihood of regional - or even global - war. In addition, deglobalization (Goldberg and Reed (2023)) and in particular a strengthening of economic ties inside military alliances - such as that observed between Russia, China, Iran, and North Korea -, together with weakening ties between alliances - such as between NATO and Russia or China - leads to a higher likelihood of conflict between military blocks, as Martin et al. (2008) forcefully argued.\footnote{Martin et al. (2008) argue that a rise in multilateral trade and a fall in bilateral trade in the case of two countries raises the probability of conflict between them, due to a reduction in the opportunity cost of war.}

Under these circumstances, military buildups become necessary to deter and potentially fight back foreign armed interventions. The consequences of war preparations have wide-reaching implications for economic policy in general, and climate policy in particular.

This paper aims to uncover, measure, and understand the effects of militarization on climate change mitigation. In particular, I investigate two main channels. First, military buildups can raise greenhouse gas emissions directly, through increased energy and fuel use and a larger volume of military procurements. Armed forces need more energy and fuel to run their military bases, planes, bombers, tanks, etc. at a higher intensity, which increases energy use. They also need more of these weapons, and defence manufacturing is an energy-intensive part of the economy, but especially its inputs, such as primary metals, rubber, or chemicals are very energy-intensive. Hence, increasing demand for these goods can drive up the emission intensity of an economy and consequently total emissions.

Second, military buildups can impede the green transition through indirect economic effects. Such a demand shock can lead to a crowd-out in green investment and innovation. Defence and fossil fuel industries will require large amounts of investment to expand capacity to the level where it can supply a substantially more active military. This can lead to a scarcity of investment and innovation funds for climate change mitigation and adoption, delaying and potentially disrupting the green transition.

I find evidence for both of these mechanisms. In my empirical analysis, I use country-year panel data, extract military spending shocks using the method proposed by Hamilton (2018), and estimate the dynamic effect of such a shock on emissions and emission intensity using panel local projections. The results indicate that total emissions rise by 0.9-2\%, while emission intensity rises by about 1\% in response to a percentage point rise in the military spending share.\footnote{Both the empirical analysis and the model results suggest that a rise in emission intensity, and not simply a rise in output due to a positive spending multiplier, is the main driving force behind these results.} The result depends on how "dirty" is the given economy - that is, how emission-intensive the production structure is.

I also find evidence that military buildups reduce green innovation. The empirical analysis indicates that climate change mitigation and adaptation patents fall by 10-25\% following a military expenditure shock, which can happen either due to a crowding-out effect or due to a general reduction in innovation. 

I then build a dynamic production network model which can capture inter-industry substitution possibilities, shock propagation, and capacity expansion dynamics (through an investment network). I calibrate the model to the United States, with data from 2017. The model indicates that a percentage point rise in the military spending share would increase total emissions by between 0.36\% and 1.81\%, and emission intensity by between 0.22\% and 1.5\%.

In addition, the model indicates that investment in the utilities sector might experience a slump in response to a temporary military spending shock, which suggests that investment in new renewable energy sources will most likely also suffer.

The result that the risk of military conflict can worsen climate change is important on its own because it shows that there is a "green peace dividend" - an additional cost of conflict. In section 5, I show that the measured effects can significantly raise climate damages and increase temperatures. The results from the model indicate that if the military spending share was doubled in the US in 2017 (when it was 3.3\% of GDP), damages due to higher emissions would equal between 0.07-2.6\% of GDP per year, depending on the social cost of carbon and the composition of the military spending shock. This calculation does not take into account potential effects on green innovation and investment.

However, we also know that climate change can increase the likelihood of conflicts (see, e.g., Burke, Hsiang, and Miguel (2015) for a good summary of this literature). Hence, a positive feedback loop between conflict and climate change can increase the destructiveness of both. I leave the evaluation of this feedback loop to future work.

Policymakers can counteract these negative effects using either a "stick" or a "carrot" approach, or a combination of these. The former would entail a rise in carbon prices, reducing total GHG emissions. Unfortunately, it would most likely hit political and social constraints: military buildups and wars require welfare sacrifices in general, and carbon prices would most likely pile upon these pressures. The "carrot", investment and innovation subsidies, would likely be more popular, and could be harmonised with defence goals such as energy security.

\subsection*{Literature review}

This paper contributes to the literature on the links between military activity and environmental damages. It is closest to the sub-stream looking at the causal effects of military expenditures on greenhouse gas emissions. Bradford and Stoner (2017), Clark et al. (2010), Jorgenson and Clark (2015), Dong et al. (2024), and Bildirici (2016) conduct empirical analyses of the effect of military expenditures on emissions or emission intensity. They generally find positive effects, although Bradford and Stoner (2017) find a relatively small impact of the military expenditure share on emission intensity. 

Most of these papers ignore the non-stationarity of the data (although Bildirici (2016) tests and estimates the cointegrating relationship between military expenditures and emissions) and run simple panel regressions. Compared to them, I extract military spending shocks and calculate the dynamic effects of such a shock on emissions and emission intensity. In addition, I also use a dynamic production network model to estimate these causal effects, as well as to uncover the underlying mechanisms. I also analyse the effect of such a shock on other long-run economic consequences, such as green investments or innovation - I do not restrict my focus on GHG emissions.

This paper is also related to work on the carbon footprint of militaries. Crawford (2019, 2022) finds that the Pentagon is the "world’s	largest	institutional	user of petroleum and correspondingly, the single largest institutional producer of greenhouse gases in the world". Parkinson and Cottrell (2021, 2022) estimate EU and global military GHG emissions, while Parkinson (2020) estimates the UK military sector's environmental impact. Sparrevik and Utstol (2020) measure and analyse the Norwegian defence sector's carbon footprint. Compared to these papers, I focus on the causal effect of military activity on emissions and the green transition instead of measuring the current level of emissions linked to the military.

My paper also contributes to the literature on the environmental effects of wars and conflict. It is closest to two papers in particular. Neimark et al. (2024) and de Klerk et al. (2023) analyse the environmental costs of the Israel-Hamas war and the Ukraine war, respectively. Compared to them, I do not focus specifically on wars, and I also look at the longer-term effects of military buildups on the green transition.

Ramey and Shapiro (1998), Ramey (2011), Ilzetzki et al. (2013), Auerbach and Gorodnichenko (2012), Hall (1980), Hall (2009), Barro (1981), Barro and Redlick (2011), Gordon and Krenn (2010), Nakamura and Steinsson (2014), and Miyamoto et al. (2019) exploit military spending shocks to measure the effect of government spending on different economic outcomes - however, they do not analyse the link between military spending and emissions. Couttenier et al. (2022) and Korovkin and Makarin (2022) use production network models to estimate the economic effects of wars, in the context of the Maoist insurgency in India and the Ukraine war, respectively. I use methods similar to these papers to understand the effect of military activity, but I focus on climate-relevant impacts. 

The paper proceeds in the following manner. I first discuss the empirical analysis in section 2, followed by the economic model in section 3. I sum up and interpret the results in section 4, where I discuss the Green Peace Dividend and potential policy implications, after which I conclude.

\section{Empirical analysis}

In my empirical analysis, I aim to uncover how militarization affects greenhouse gas emissions, as well as green innovation. I proxy militarization with the military expenditure share of GDP.

\subsection{Method and Data}

\subsubsection{Data}

I use annual GHG emissions data from EDGARv8.0. I also use data from the Correlates of War (National Material Capabilities version 6.0) for energy consumption in tons of coal equivalent, as well as for military expenditures, military personnel, crude steel production, population, and urban population. National accounts data is from the Penn World Tables (version 10.01). I'm using data from V-Dem v13 measuring how close a given country is to the "ideal of electoral democracy" (Coppedge et al, 2023). I use OECD data for total and environmental patents.

The data covers the period 1970-2016, except for the data on patents, which covers 1978-2016. In my main sample, I include 20 NATO countries: Belgium, Bulgaria, Canada, Denmark, Finland, France, Germany, Greece, Hungary, Italy, the Netherlands, Norway, Poland, Portugal, Romania, Spain, Sweden, Turkey, the United Kingdom, and the United States. These are countries that are likely to have good data for the variables of interest in the period analysed. They also yield a balanced panel in my analysis (see below). In addition, the effect of military buildups in these countries - NATO members - is of special interest following the Russian invasion of Ukraine. Many of them are also countries that have some arms industry, which raises the domestic absorption of the military budget (I discuss below why this is important). 

To check the robustness of my results, I also consider an extended sample of 38 countries, covering almost 80\% of 2024 global PPP GDP\footnote{Based on IMF data: \url{https://www.imf.org/external/datamapper/PPPSH@WEO/OEMDC/ADVEC/WEOWORLD?year=2024}} (see table 1).\footnote{Two countries have data only from 1990 on: Russia and the Czech Republic. I included these for the importance of their arms industry (currently or historically), but removing them does not change the results.} See table 1 for the list of countries in this extended sample.

\subsubsection{Method}

\begin{figure}[h!]
    \centering
    \includegraphics[width=0.5\textwidth]{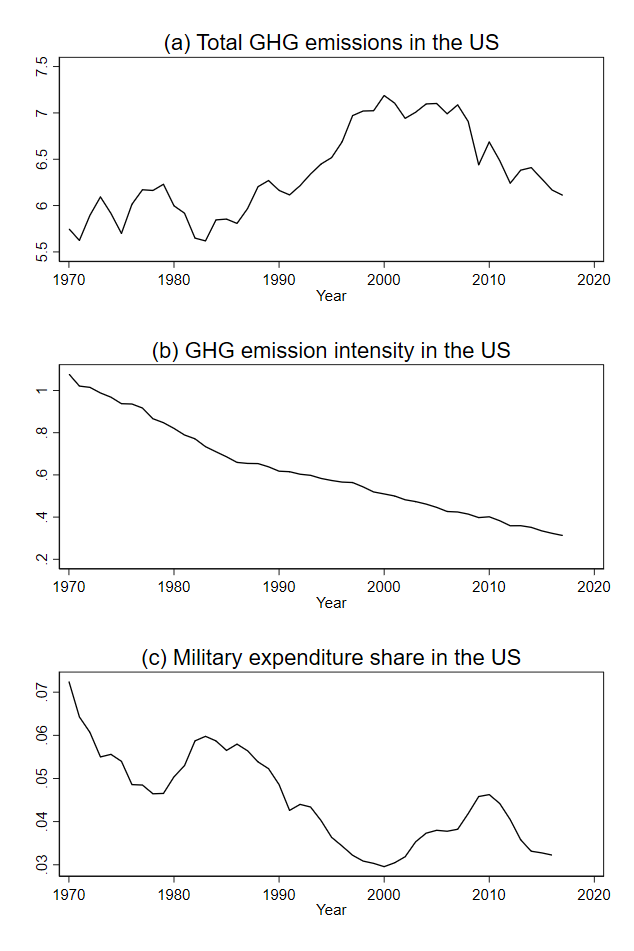}
    \caption{Trends in variables of interest}
    \label{fig:emp1}
    \caption*{Note: the figure shows the evolution of total GHG emissions, emission intensity, and the military expenditure share of GDP in the US.}
\end{figure}

The main variables of interest - total GHG emissions, emission intensity (defined as GHG emissions per unit of real product), and the military expenditure share - follow clear time trends. Figure \ref{fig:emp1} shows that, in the United States, both emission intensity and the military expenditure share trended down between 1970 and 2016, while total emissions first followed an upward trend, which inverted during the mid-2000s. Ignoring this could lead to estimating spurious relationships. 

To extract the causal effect of military spending on the outcomes of interest, I follow the approach in Bilal and K\"anzig (2024) and first extract military spending shocks using a Hamilton (2018) filter, after which I estimate the dynamic effect of such a shock on the outcomes of interest using panel local projections (Jord\`a (2005), Jord\`a et al. (2020)). I assume that these shocks are exogenous, as sudden changes in defence spending are most likely to follow strategic, and not economic, considerations. It is common in the literature to assume that military spending shocks occur due to geopolitical events or strategic considerations, and not due to endogenous economic considerations. For example, Miyamoto et al. (2019) assume that "changes in military spending are exogenous to economic conditions". A similar argument can be found in Hall (1980).

In particular, I first use the method proposed in Hamilton (2018) to extract shocks. That is, I run the regression:
\begin{gather}
    M_{i,t+h} = \alpha_{i} + \beta_{i,1}M_{i,t} + ... +  \beta_{i,l+1}M_{i,t-l} + \epsilon_{i,t+h}
\end{gather}

For some horizon $h$ and lag order $l$, separately for each country $i$ in the sample. $M$ denotes the military spending share of GDP. Following this, I define the residuals of the regression as the military spending shocks: $\hat{M}^{shock}_{i,t} = \hat{\epsilon}_{i,t}$. In practice, I use $h = 2$ and $l = 2$, but my results are robust to setting $h = 1$, that is, calculating the one-period-ahead forecast error. As an example, the military spending shocks in the US can be seen below, on figure \ref{fig:emp2}.

\begin{figure}[h!]
    \centering
    \includegraphics[width=0.5\textwidth]{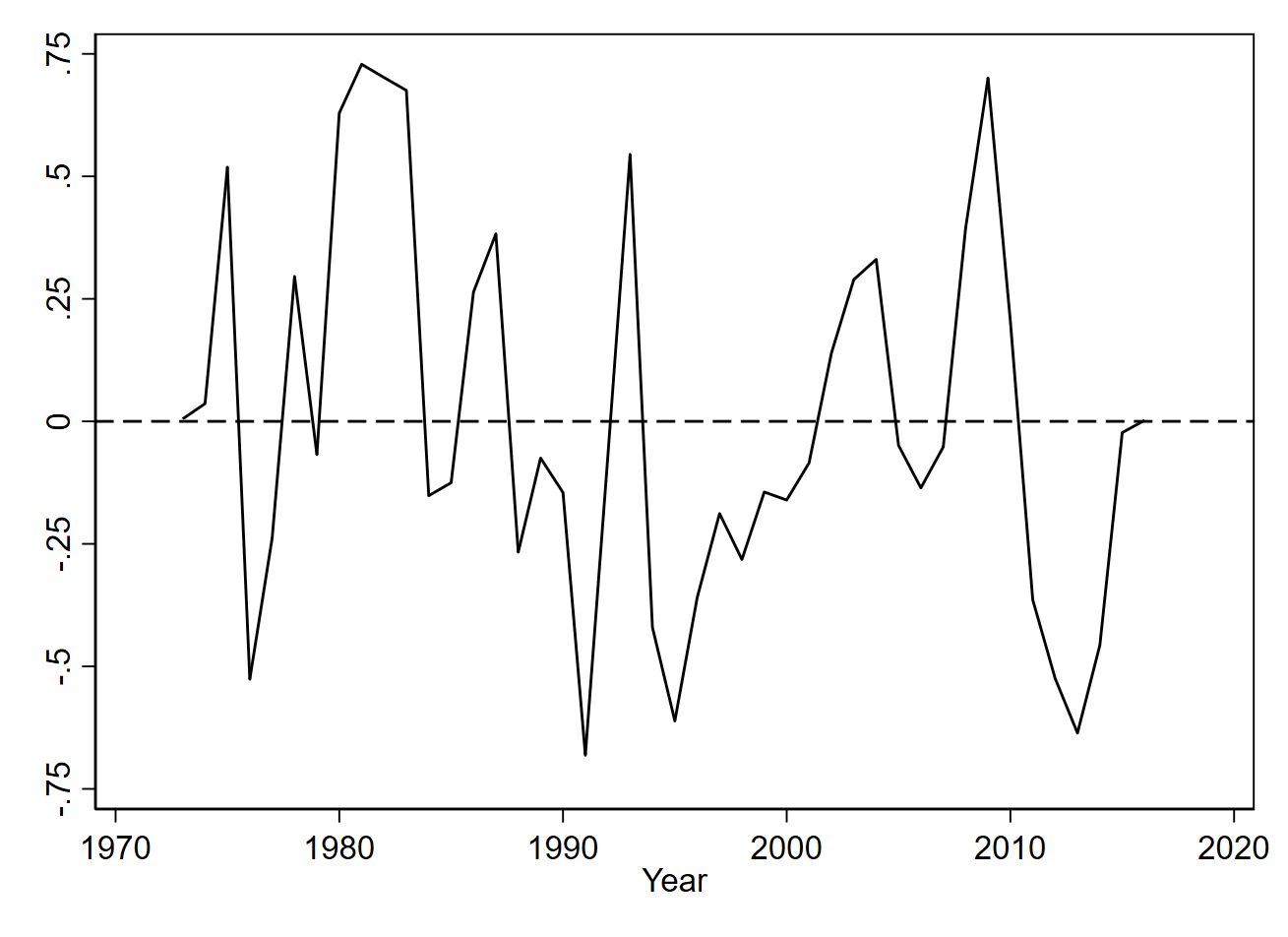}
    \caption{Military spending shocks}
    \label{fig:emp2}
    \caption*{Note: the figure shows the military spending shocks in the US, extracted using the Hamilton (2018) method, expressed in percentage deviations from the trend.}
\end{figure}

Following this, I use these shock series to estimate the IRF of a military spending shock, equivalent to increasing the military spending share by one percentage point. In particular, I run the following local projections:
\begin{gather}
    y_{i,t+h} - y_{i,t-1} = \alpha_{i}^{h} + \delta_{t}^{h} + \beta_{h}\hat{M}^{shock}_{i,t} + \theta_{1}^{h'} X_{i,t-1} + \sum_{j = 1}^{l}\theta_{2,j}^{h}\Delta y_{i,t-j} + u_{i,t+h}^{h}
\end{gather}

Where $y$ denotes either (log) emissions or (log) emission intensity, $\alpha$ is a country fixed effect, $\delta$ is a year fixed effect, $X_{i,t-1}$ is a vector of lagged controls for country i in year t, and $u$ is the regression error term. \footnote{The control variables include the first differences of logged GDP per capita, exports and imports, and a democracy variable (in levels), measuring the distance of a given country's institutions in a given year from the "ideal of electoral democracy" (Coppedge et al, 2023). I also add a second lag of GDP per capita.} I also add $l$ lags of the first difference of the outcome variable to account for potential autocorrelation in the data - in practice, I set $l = 2$. The IRF of the military spending shock is given by $IRF_{M} = \{\beta_{h}\}_{h = 0}^{T}$, where I calculate the IRF up to horizon $T = 15$. \footnote{The interpretation of $\beta_{h}$ is the following: if we increase the military spending share by 1 p.p., what's the percentage change in the outcome relative to what it would have been without the shock, at horizon h?} I use Driscoll and Kraay (1998) standard errors for inference, with two lags. In my IRFs, I show 68\% confidence bands - as Ramey (2011) points out, 68\% confidence intervals are common in the government spending multiplier literature.

\subsection{Results}

First, I will be looking at a military spending share shock's effect on emissions, after which I will look at its effects on green patenting.

\subsection*{Emissions}

\begin{figure}[h!]
    \centering
    \includegraphics[width=0.5\textwidth]{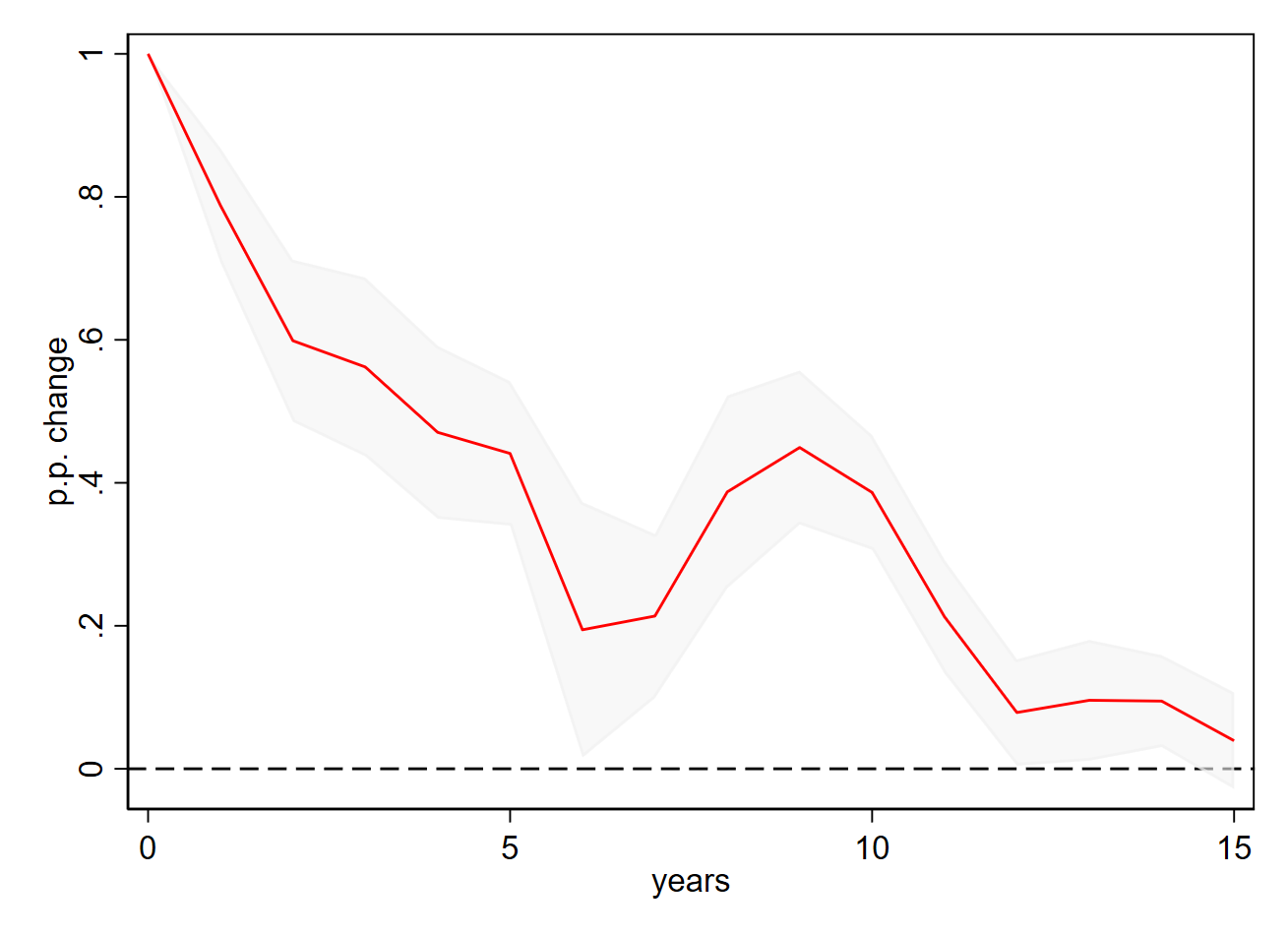}
    \caption{IRF - military expenditure share}
    \label{fig:emp3}
    \caption*{Note: the figure shows the IRF of the military spending share in response to a military spending shock corresponding to a 1 p.p. rise in the military spending share. The vertical axis shows percent changes. The dark grey area marks the 68\% confidence interval.}
\end{figure}

The first point to note is that a military spending shock has a very persistent effect on the military expenditure share. Figure \ref{fig:emp3} shows that such a shock leads to the spending share being much above trend even 10 years after the shock, slowly returning to zero. \footnote{This IRF was estimated with a local projection defined by equation 2, where $y$ denotes the military expenditure share.} Therefore, such a shock is likely to have long-lasting economic effects.

\begin{figure}[h!]
    \centering
    \includegraphics[width=0.8\textwidth]{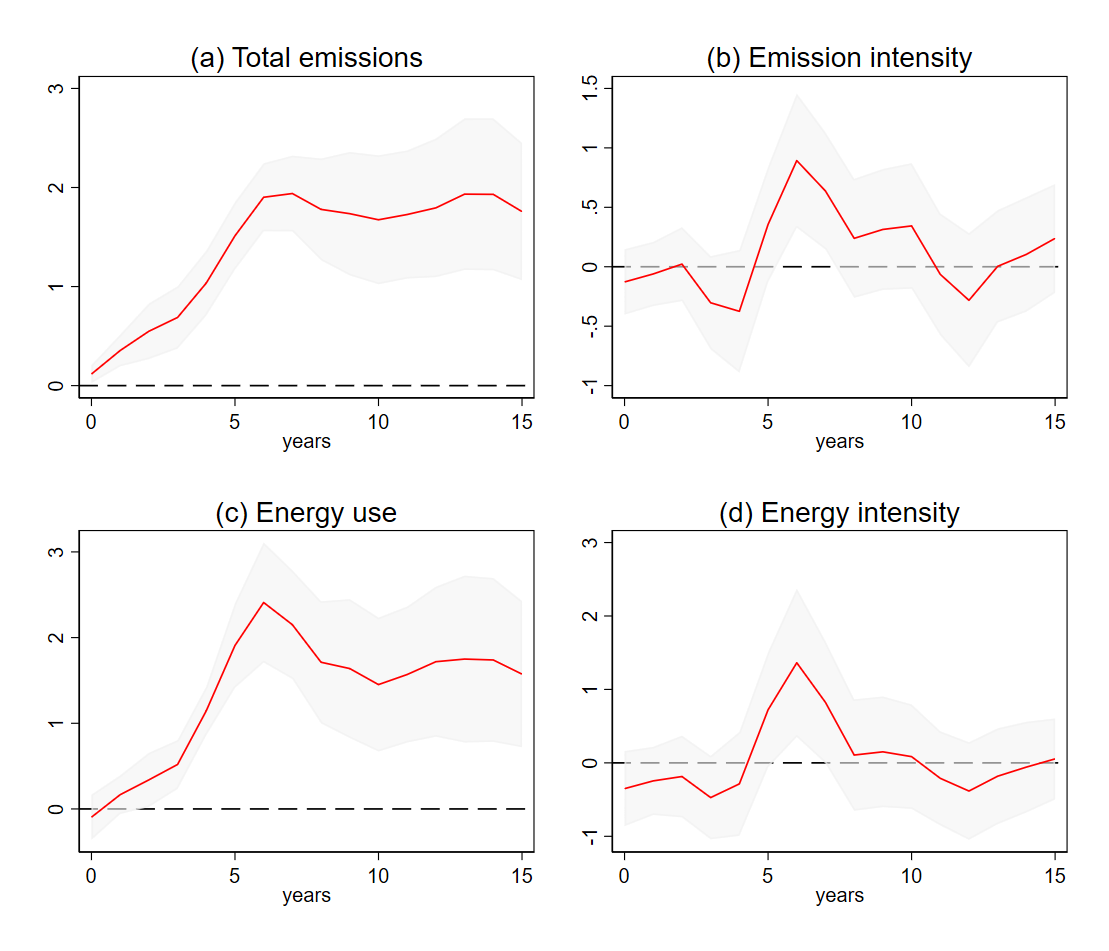}
    \caption{Baseline results - emissions and energy}
    \label{fig:emp4}
    \caption*{Note: the figure shows the IRFs of total emissions, energy use, emission intensity, and energy intensity in response to a military spending shock corresponding to a 1 p.p. rise in the military spending share. The outcome is in terms of percentage changes. The dark grey area marks the 68\% confidence interval.}
\end{figure}

Figure \ref{fig:emp4} shows the main results for the baseline, NATO sample.  Both total emissions and emission intensity (panels (a) and (b)) peak approximately 6 years after the shock - emissions rise by 1.94\%, while emission intensity rises by 0.9\% in response to an increase in the military spending share of one percentage point. Two potential reasons why emission intensity increases, as discussed above, are, first, that the armed forces start using more fuel to run their weapons and bases, increasing the fuel use of the economy, and second, that the military industries are more emission-intensive than the rest of the economy, which means that a relative expansion of these industries raises the aggregate emission intensity of the economy. 

\begin{figure}[h!]
    \centering
    \includegraphics[width=0.5\textwidth]{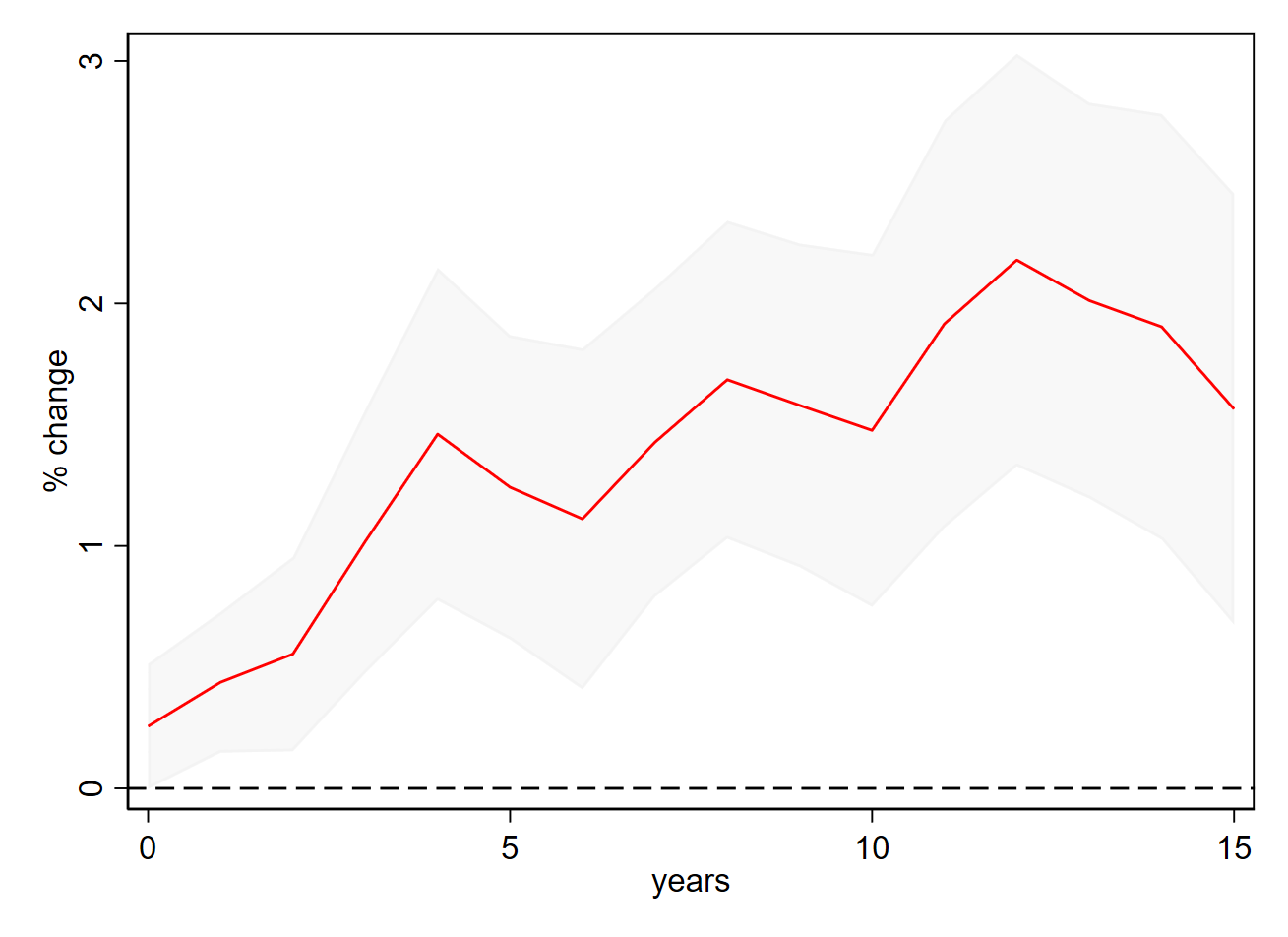}
    \caption{Real GDP}
    \label{fig:emp5}
    \caption*{Note: the figure shows the IRF of real GDP in response to a military spending shock corresponding to a 1 p.p. rise in the military spending share. The outcome is in terms of percentage changes. The dark grey area marks the 68\% confidence interval.}
\end{figure}

Total emissions rise both because emission intensity increases and because output rises.\footnote{Note that, by definition: $$Emissions = RealGDP \cdot EmissionIntensity$$ So that: $$d\ln(Emissions) = d\ln(RealGDP) + d\ln(EmissionIntensity)$$} Figure \ref{fig:emp5} shows that output also rises in response to such a shock, by 1.11\% after 4 years, which falls to approximately 0.67\% after 6 years. This result is in line with the evidence in Ramey (2011), Auerbach and Gorodnichenko (2012), and Ilzetzki et al. (2013) concerning industrial countries (my main sample consists almost completely of industrialized countries).\footnote{A CfM survey in July 2024 by Ilzetzki and Jain also found that most experts believed an increase in defence spending will boost economic growth in the EU over a 5-year horizon. \url{https://www.cfmsurvey.org/survey-2024-07}}

Emission intensity seems to increase significantly only in the horizon of 6-7 years. The bulk of the military activity may concentrate on this horizon after the shock - however, more evidence is needed to understand the dynamics of emission intensity following military buildups. Nevertheless, the rise in emission intensity makes up almost half of the peak increase in total emissions, suggesting that the rise in emissions is not simply due to a positive government spending multiplier driving up output.

\begin{figure}[h!]
    \centering
    \includegraphics[width=0.5\textwidth]{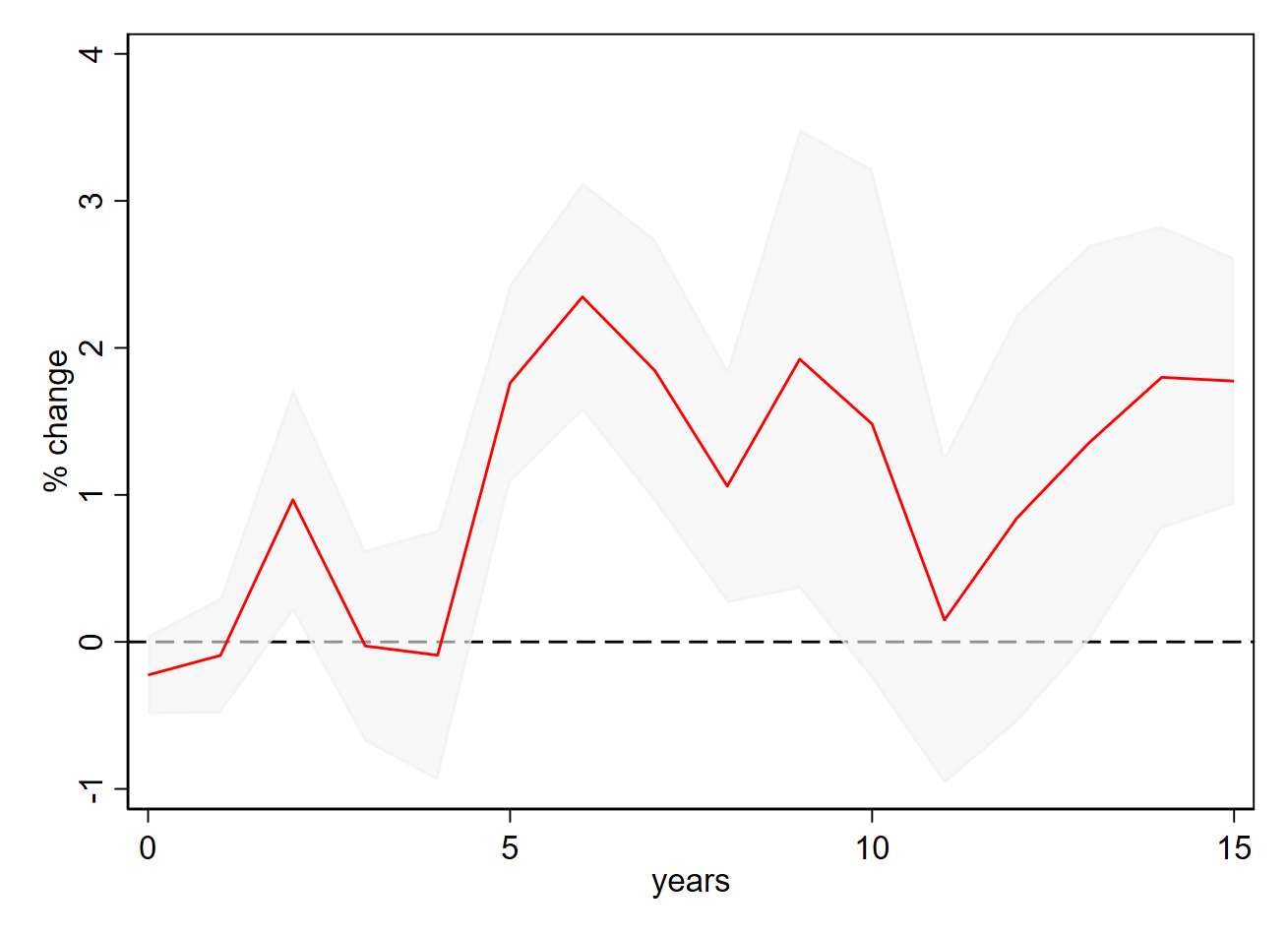}
    \caption{Crude steel production}
    \label{fig:emp6}
    \caption*{Note: the figure shows the IRF of crude steel production in response to a military spending shock corresponding to a 1 p.p. rise in the military spending share. The outcome is in terms of percentage changes. The dark grey area marks the 68\% confidence interval.}
\end{figure}

I will discuss the mechanism behind the effect of military spending on emissions in section 3. However, figure \ref{fig:emp6} shows that a defence spending shock increases crude steel production in the baseline sample countries by more than 2\% after 6 years, which indicates that heavy manufacturing experiences a significant boost (even relative to real GDP, which rises by approximately 1\% at this horizon). This provides evidence that military spending shocks target energy-intensive industries, such as steel manufacturing.

Note that contemporary effects are virtually zero. A potential explanation for why the effect peaks much later than the year of the shock is that military buildups require time to materialize: weapon procurements can take years, and yet more time is needed for new weapons to be operationalized and new soldiers to be trained.\footnote{This can, of course, happen in much shorter time periods, depending on the urgency of the military situation. During wartime, major industrial powers are capable of producing a very significant amount of mat\`eriel in a very short time - an extreme example of this can be found in the case of the Second World War and the performance of the formidable military industries of the United States, the Soviet Union and Nazi Germany. See, for example, Ilzetzki (2024). No such major events are included in my sample.} Military buildups may require large-scale investments into military industries and their upstream suppliers, which can take time to materialise and become functional.

Instead of emissions, one can also use energy use as an outcome variable to cross-check the results - see panels (c) and (d) on figure \ref{fig:emp4}. The results are very similar if we look at total energy use and energy intensity. Note that energy use does not necessarily translate one-to-one into emissions - the latter depends on the energy composition as well. Using data on the energy mix\footnote{Energy Institute - Statistical Review of World Energy (2024) – with major processing by Our World in Data. “Biofuels consumption” [dataset]. Energy Institute, “Statistical Review of World Energy” [original data].}, I find that the share of fossil fuels out of total energy consumption (in TWh) increases by 0.3 percentage points at the peak, driven mainly by higher natural gas consumption, suggesting that energy use becomes slightly "dirtier".

Note that there is an effect which can reduce the measured impact on both emissions and emission intensity. Usually, part of the military budget is spent in foreign countries - both weapons (through arms trade) and energy could be bought externally. Indeed, a large share of EU defence procurement spending goes to non-EU suppliers - between June 2022 and June 2023, 78\% of EU defence procurement spending went to non-EU suppliers (see the Draghi report (2024)). This means that part of the rise in emissions and emission intensity due to military buildups could show up in other countries' data, which suggests that my results are conservative estimates of the true effect.

\subsection*{Extended sample}

To check the generality of the previous results, I rerun the previous local projections with the extended sample of 38 countries, mentioned before. The results are much weaker than those for the main sample: total emissions virtually do not change. There might be several reasons behind this. First, the domestic absorption of the additional countries in the sample might be lower than that of the NATO sample - countries such as Ireland likely do not source most of their military procurements domestically. 

Alternatively, the countries in the sample with high emission intensity may emit relatively more in response to a military spending shock - for example, because their manufacturing sector is more emission-intensive relative to the rest of the economy. Energy could be cheaper and more easily accessible in these countries relative to other factors of production, which might mean that industries rely disproportionately on energy-intensive technologies to satisfy rising weapon demand during military buildups. 

I check whether this is indeed the case following the method proposed by Miyamoto et al. (2019). First, I separate countries into high-emission intensity and low-emission-intensity categories by calculating their average emission intensity through time and assigning those above the median to the high-emission-intensity category, and vice versa. The countries in these two categories can be seen in table 1 below.

\begin{center}
\resizebox{4in}{!}{
\begin{tabular}{cc}
 High emission intensity countries & Low-emission-intensity countries \\ \hline
 Argentina & Austria \\
 Australia & Chile \\
 Belgium & Denmark \\
 Brazil & France \\
 Bulgaria & Germany \\
 Canada & Greece \\
 China & Israel \\
 Czech Republic & Italy \\
 Finland & Japan \\
 Hungary & Mexico \\
 India & Netherlands \\
 Ireland & Norway \\
 South Korea & Portugal \\
 New Zealand & Spain \\
 Poland & Sweden \\
 Romania & Switzerland \\
 Russia & Taiwan \\
 South Africa & Turkey \\
 United States & United Kingdom \\\hline
\multicolumn{2}{c}{Table 1. High- and low-emission intensity countries in the extended sample} \\
\end{tabular}}
\end{center}

Then, I run the following local projections:

\begin{gather*}
    y_{i,t+h} - y_{i,t-1} = \mathbf{I}_{i,H}(\alpha_{i}^{h} + \delta_{t,H}^{h} + \beta_{h,H}\hat{M}^{shock}_{i,t} + \theta_{1,H}^{h'} X_{i,t-1} + \sum_{j = 1}^{l}\theta_{2,j,H}^{h}\Delta y_{i,t-j})\\
    + \mathbf{I}_{i,L}(\alpha_{i}^{h} + \delta_{t,L}^{h} + \beta_{h,L}\hat{M}^{shock}_{i,t} + \theta_{1,L}^{h'} X_{i,t-1} + \sum_{j = 1}^{l}\theta_{2,j,L}^{h}\Delta y_{i,t-j}) + u_{i,t+h}^{h}
\end{gather*}

\begin{figure}[h!]
    \centering
    \includegraphics[width=0.5\textwidth]{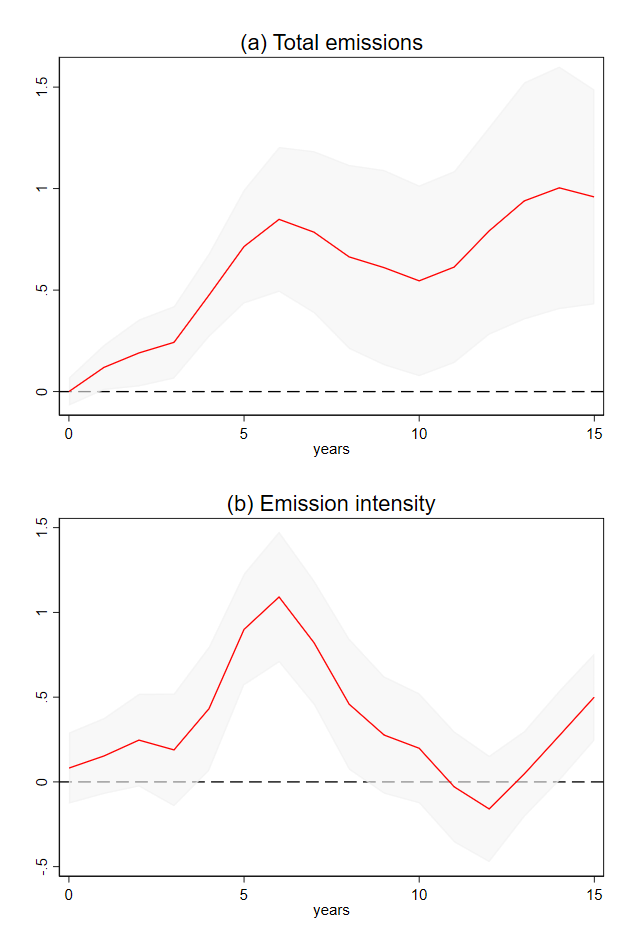}
    \caption{High emission intensity countries - results}
    \label{fig:emp7}
    \caption*{Note: the figure shows the IRFs of total emissions and emission intensity in response to a military spending shock corresponding to a 1 p.p. rise in the military spending share, in high emission intensity countries. The outcome is in terms of percentage changes. The dark grey area marks the 68\% confidence interval.}
\end{figure}

Where $\mathbf{I}_{i,H}$ is equal to one if country i is classified as high emission intensity and zero if not, while $\mathbf{I}_{i,L}$ is equal to one if country i is classified as low emission intensity and zero if not. $\{\beta_{h,H}\}_{h = 0}^{T}$ and $\{\beta_{h,L}\}_{h = 0}^{T}$ are, then, the IRFs of the high and low emission intensity countries. The high-emission-intensity IRFs can be seen on figure \ref{fig:emp7}, while the low-emission-intensity IRFs are on figure \ref{fig:emp9}.

\begin{figure}[h!]
    \centering
    \includegraphics[width=0.5\textwidth]{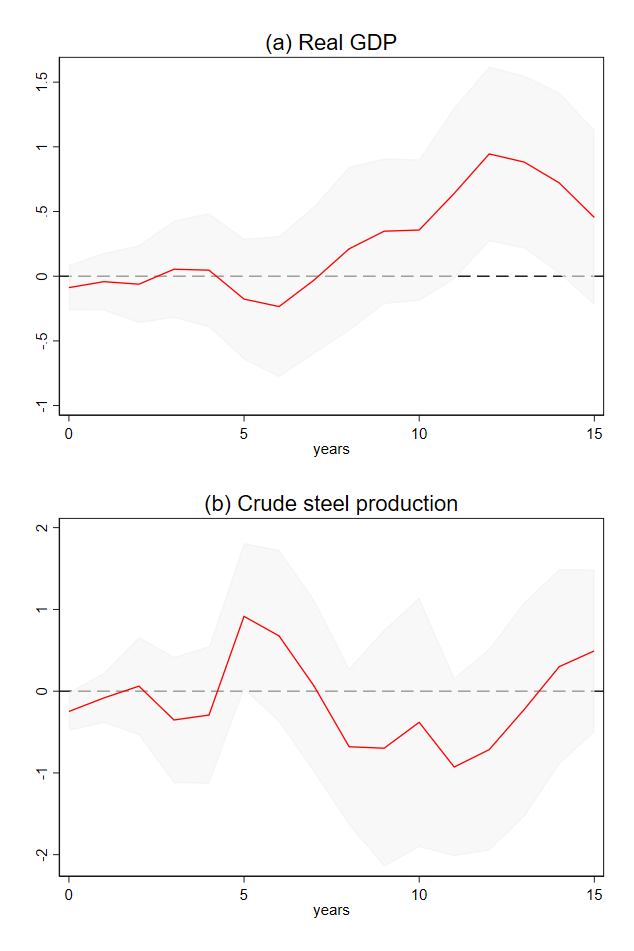}
    \caption{High emission intensity countries - output and steel production}
    \label{fig:emp8}
    \caption*{Note: the figure shows the IRFs of real GDP and crude steel production in response to a military spending shock corresponding to a 1 p.p. rise in the military spending share, in high emission intensity countries. The outcome is in terms of percentage changes. The dark grey area marks the 68\% confidence interval.}
\end{figure}

In high-emission-intensity countries, emissions rise by approximately 0.9\% 6 years after the shock, while emission intensity rises to a peak of more than 1\%. This indicates that, in this sample, military spending shocks drive emissions up solely by making the economy "dirtier": figure \ref{fig:emp8} shows that real GDP does not change significantly following a shock and that crude steel production rises by 1\% at a horizon of 5 years.

In low-emission-intensity countries, however, a military spending shock significantly reduces total emissions - by 3\% after 3 years (see figure \ref{fig:emp9}). It similarly reduces emission intensity, by 2\%. Figure \ref{fig:emp10} suggests that both output and crude steel production fall in this sample, which could suggest that, instead of the production structure being "cleaner" in this sample, the results could be driven by low domestic absorption (redirecting demand from domestically produced goods to foreign-produced weapons).

\begin{figure}[h!]
    \centering
    \includegraphics[width=0.5\textwidth]{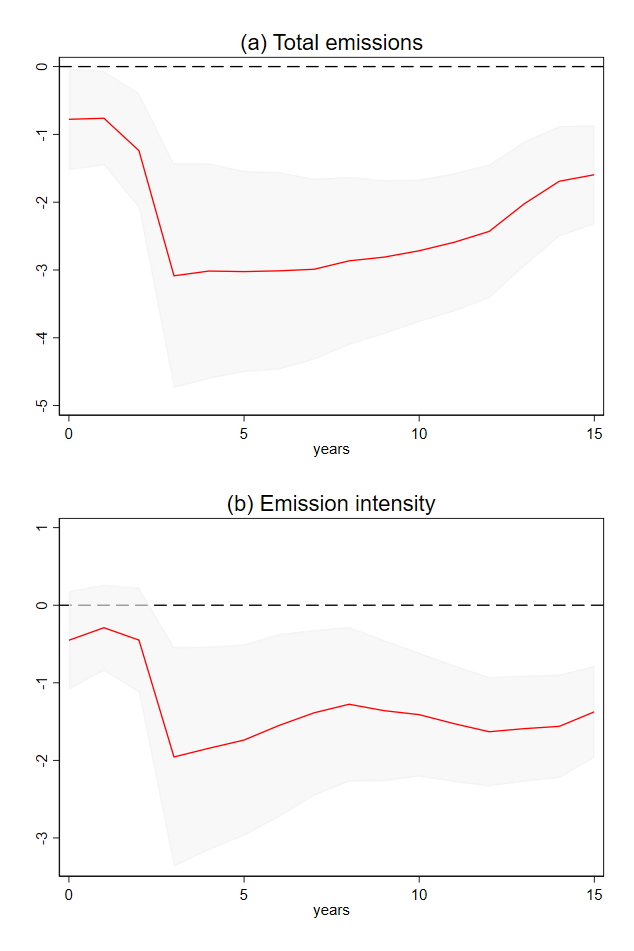}
    \caption{Low emission intensity countries - results}
    \label{fig:emp9}
    \caption*{Note: the figure shows the IRFs of total emissions and emission intensity in response to a military spending shock corresponding to a 1 p.p. rise in the military spending share, in low emission intensity countries. The outcome is in terms of percentage changes. The dark grey area marks the 68\% confidence interval.}
\end{figure}

\begin{figure}[h!]
    \centering
    \includegraphics[width=0.5\textwidth]{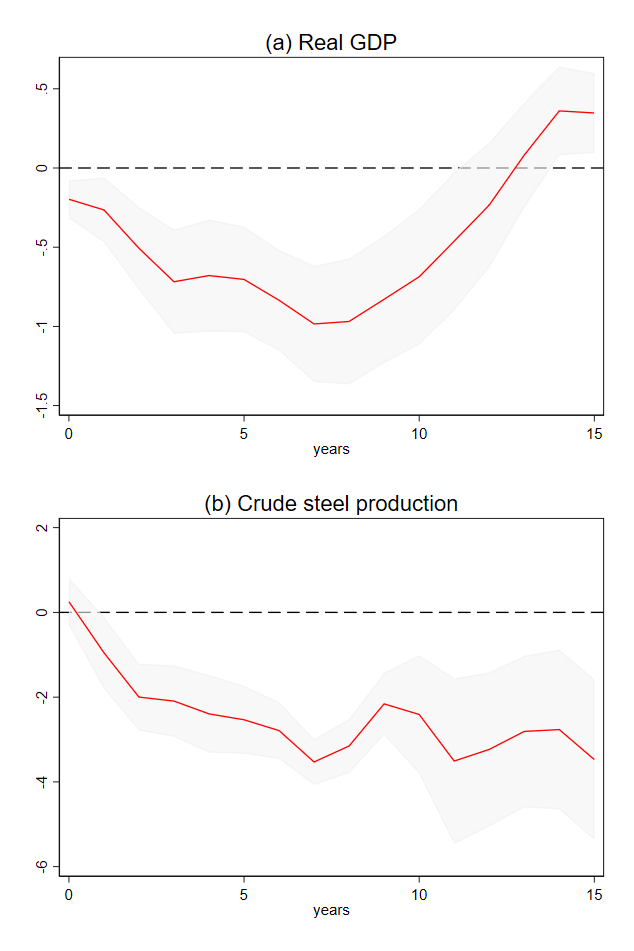}
    \caption{Low emission intensity countries - output and steel production}
    \label{fig:emp10}
    \caption*{Note: the figure shows the IRFs of real GDP and crude steel production in response to a military spending shock corresponding to a 1 p.p. rise in the military spending share, in low emission intensity countries. The outcome is in terms of percentage changes. The dark grey area marks the 68\% confidence interval.}
\end{figure}

Clearly, there is evidence of a differential response to a military spending shock in these two classes of countries. Hence, the emission and energy intensity of an economy may be important in determining whether such a shock actually leads to more GHG emissions or not. However, more research is needed into the exact reasons behind this result - for example, differences may be due to higher domestic absorption of defence expenditures in high emission intensity countries.

\subsection*{Innovation}

In order to understand the effect of a military spending shock on green innovation, I look at the effects on patents linked to climate change mitigation and adaptation using the local projection setup (2). However, several countries in my sample produce few or no patents targeting these areas - hence, I focus on countries in the 38-country sample that are the most innovative in this area. That is, I calculate the time series averages of green patents in each country between 1978 and 2016 and only use the countries which are above the median of these averages.\footnote{These countries are: Australia, Austria, Belgium, Canada, Chile, Denmark, Finland, France, Germany, Italy, Japan, Korea, the Netherlands, Sweden, Switzerland, Taiwan, the UK and the US.} Then, I run (2) for these countries, for the period between 1978-2016.

\begin{figure}[h!]
    \centering
    \includegraphics[width=0.55\textwidth]{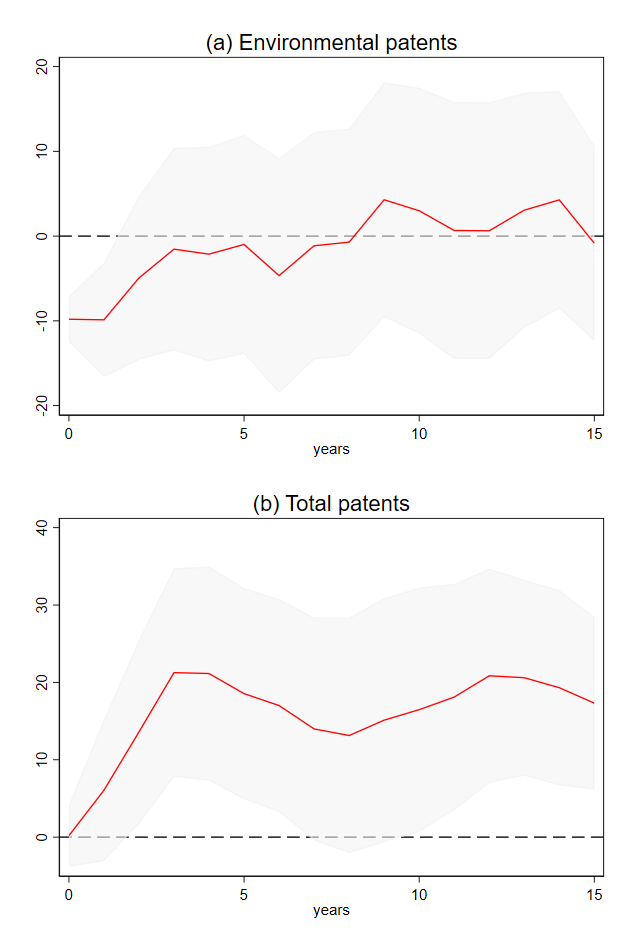}
    \caption{Patenting - green and total}
    \label{fig:emp11}
    \caption*{Note: the figure shows the IRFs of total patents and green patents in response to a military spending shock corresponding to a 1 p.p. rise in the military spending share. The outcome is in terms of percentage changes. The dark grey area marks the 68\% confidence interval.}
\end{figure}

The results can be seen on figure \ref{fig:emp11}. A military spending share shock reduces environmental patents by 10\% contemporaneously and after a year, following which it returns to the trend (panel (a)). This is not due to a general negative effect on patenting: panel (b) shows that total patents increase significantly, by more than 20\% following a defence shock. 

Total patents could increase due to two effects: first, because military R\&D boosts innovation directly, and second, because defence R\&D could "crowd-in" private innovation (see Moretti, Steinwender, and Van Reenen (2021) on the "crowd-in" effects of defence R\&D). However, the results suggest that as overall innovation receives a boost, green innovation is crowded out - potentially due to the lack of strong links between military and green research.

\begin{figure}[h!]
    \centering
    \includegraphics[width=0.55\textwidth]{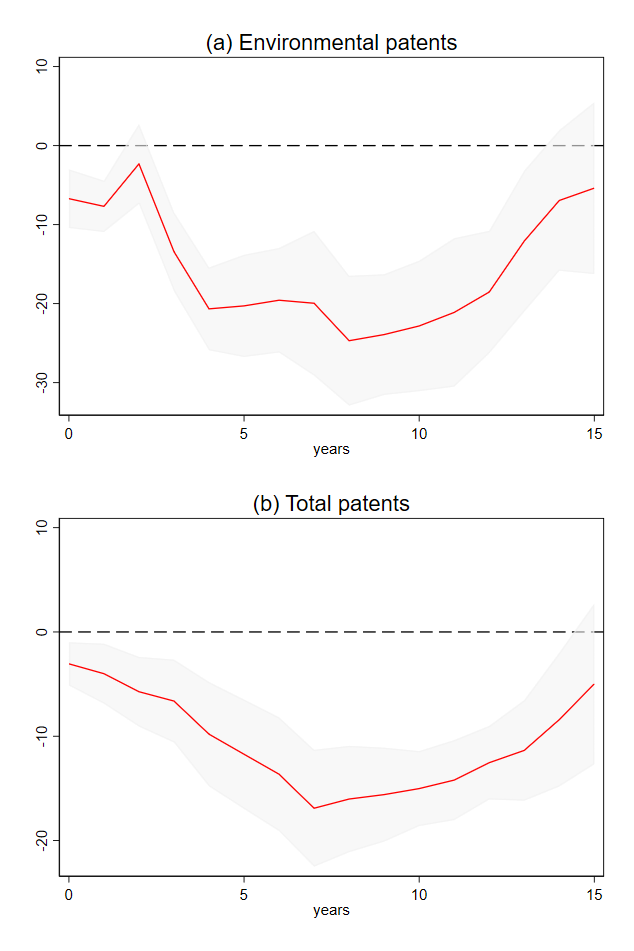}
    \caption{Patenting - green and total (NATO countries)}
    \label{fig:emp12}
    \caption*{Note: the figure shows the IRFs of total patents and green patents in response to a military spending shock corresponding to a 1 p.p. rise in the military spending share, in NATO countries. The outcome is in terms of percentage changes. The dark grey area marks the 68\% confidence interval.}
\end{figure}

I further restrict the sample to look at high-patenting NATO countries - results can be seen on figure \ref{fig:emp12}. Both environmental patenting, as well as total patents experience a sharp drop following a 1 p.p. rise in the military expenditure share. Green patenting falls relatively more than total patents (by 25\% vs. 17\%), suggesting that green innovation suffers disproportionately in NATO countries following a defence spending shock. Nevertheless, the results imply that military expenditures might have a general negative effect on innovation and not simply crowd out green patenting in this smaller sample.\footnote{Note that Moretti et al. (2021) find that defence-related R\&D expenditures increase industry-level private R\&D expenditures and, hence, total industry-level R\&D expenditures. Hence, a possible reason why total patenting falls in my smaller, NATO sample is that the identified military expenditure share shocks might contain small R\&D shares.}

As the results - especially for the larger sample of innovative countries - are quite noisy with wide confidence bands, further research is necessary to understand the effect of defence expenditure shocks on green innovation.

\subsection{Robustness checks}

\subsection*{Alternative transformations of military spending}

One concern which could arise with regard to the shock identification procedure is that the military expenditure share, $M_{t} = m_{t}/Y_{t}^{N}$ (where $m_{t}$ is nominal military expenditure at time t, and $Y_{t}^{N}$ is nominal GDP) depends on a potentially endogenous factor - nominal output - even if military expenditures are exogenous, as argued before. Concretely, it would be problematic if I were capturing recessions and not military buildups (that is, drops in the denominator instead of rises in the numerator).

In order to remove this potential source of endogeneity, I consider two alternative formulations. First, instead of using the military expenditure share, I use a Gordon-Krenn transformed military expenditure variable (Gordon and Krenn (2010)). That is, I use:
\begin{gather*}
    M^{GK}_{t} = \frac{m_{t}^{*}}{Y^{*}_{t}}
\end{gather*}
Where $m_{t}^{*}$ is nominal military spending divided by the GDP deflator, while $Y^{*}_{t}$ is a measure of "potential GDP" - in my case, I set this equal to the quadratic trend of real GDP, following Ramey (2016). Unlike actual nominal GDP, potential GDP does not vary much and does not depend on the business cycle, and hence is unlikely to be endogenous in our setting.

\begin{figure}[h!]
    \centering
    \includegraphics[width=0.8\textwidth]{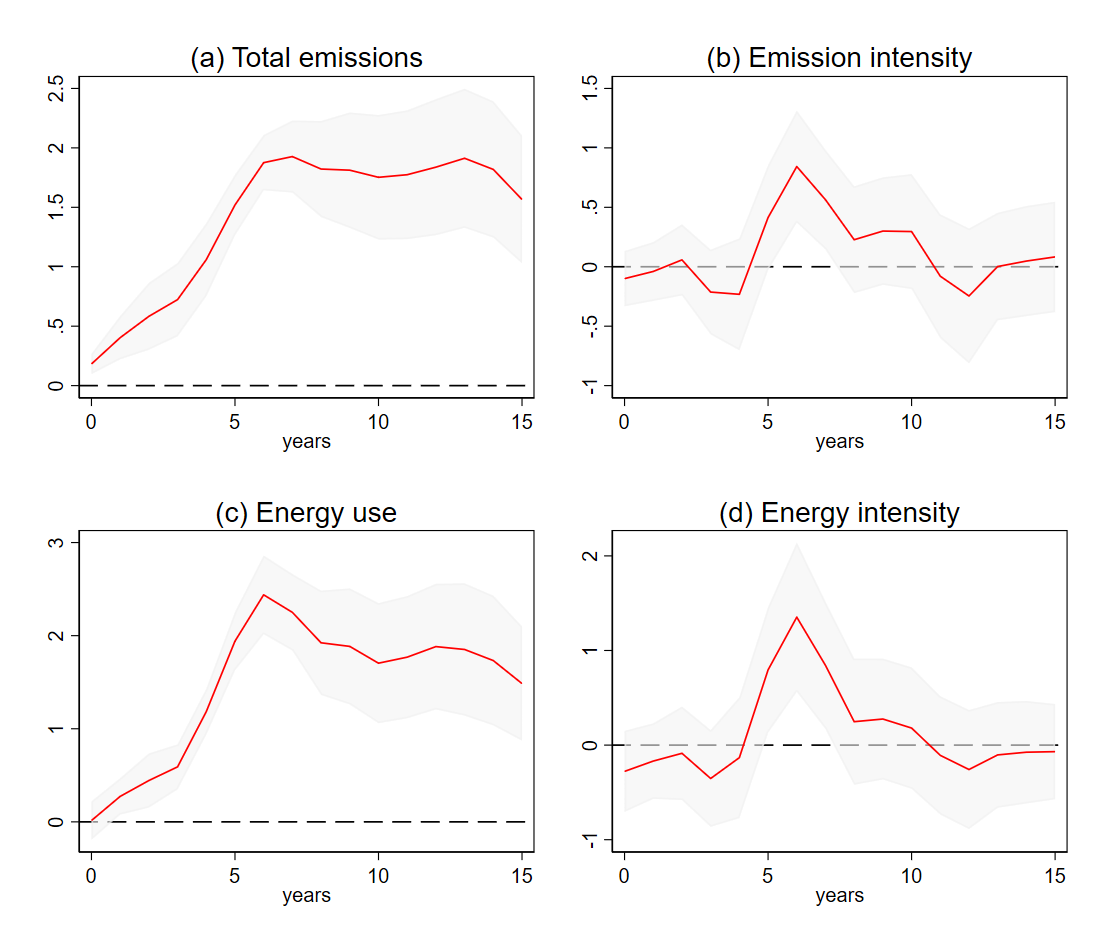}
    \caption{Gordon-Krenn transformed military expenditures - emissions and energy}
    \label{fig:emp13}
    \caption*{Note: the figure shows the IRFs of total emissions, energy use, emission intensity, and energy intensity in response to a military spending shock corresponding to a 1 p.p. rise in the Gordon-Krenn military spending share. The outcome is in terms of percentage changes. The dark grey area marks the 68\% confidence interval.}
\end{figure}

I extract shocks to the Gordon-Krenn transformed military spending share, $M^{GK}_{t}$, using the Hamilton (2018) method as before, and run the local projections with the baseline NATO sample and specification. The results can be seen on figure \ref{fig:emp13} - instead of a 1 p.p. rise in the military spending share, the shock can now be interpreted as a 1 p.p. rise in the share of real military spending in potential GDP. Clearly, the results are extremely similar to the baseline results on figure \ref{fig:emp4}.

\begin{figure}[h!]
    \centering
    \includegraphics[width=0.8\textwidth]{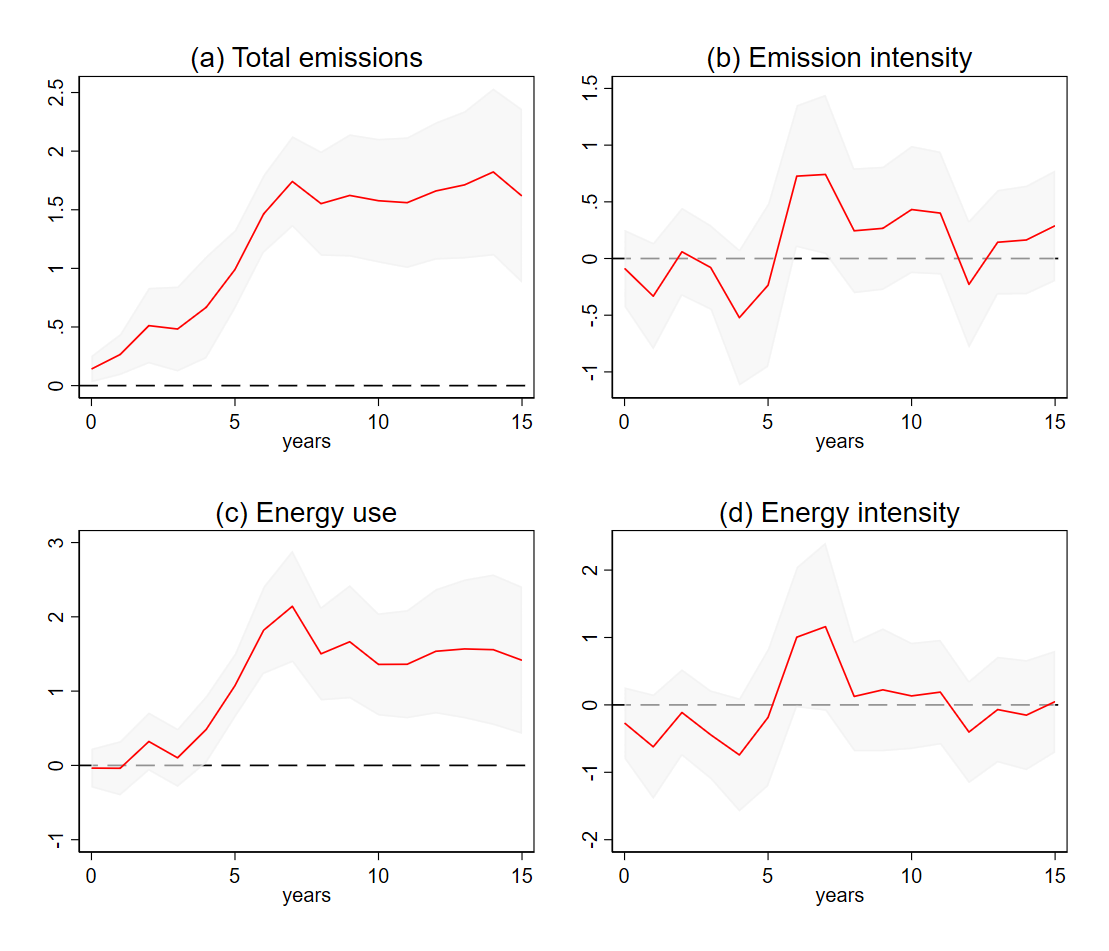}
    \caption{Hall-Barro-Redlick transformed military expenditures - emissions and energy}
    \label{fig:emp14}
    \caption*{Note: the figure shows the IRFs of total emissions, energy use, emission intensity, and energy intensity in response to a military spending shock corresponding to a 1 percent rise in military spending relative to last period's real output. The outcome is in terms of percentage changes. The dark grey area marks the 68\% confidence interval.}
\end{figure}

One can also use the Hall-Barro-Redlick transformation (following Hall (2009) and Barro and Redlick (2011)) instead of the Gordon-Krenn transformation to get rid of endogeneity. The Hall-Barro-Redlick transformed variable gives the change in real military spending as a share of year $t-1$ real GDP:
\begin{gather*}
    M^{HBR}_{t} = \frac{m_{t}^{*} - m_{t-1}^{*}}{Y^{R}_{t-1}}
\end{gather*}
Where $Y^{R}$ denotes real GDP, and $m^{*}$ is as before. $M^{HBR}_{t}$ depends only on the last period's real output, and is unrelated to the current period's military spending shock.\footnote{Note that I control for period $t-1$ changes in GDP per capita.} Then, I use $M^{HBR}_{t}$ directly as a shock in my local projection formulation. The results can be seen on figure \ref{fig:emp14}. Now, the shock should be interpreted as a 1 percent rise in real military spending relative to the last period's real output. Again, the results are virtually unchanged compared to the baseline results on figure \ref{fig:emp4}.

\subsection*{Strategic spillover effects}

If a given country raises its military spending share due to an exogenous shock shift in policy, other countries might react by changing their own military expenditures. For example, strategic allies might reduce their own expenditures because they free ride on their allies' military capabilities - or, alternatively, they might increase their own expenditures because they need to help their allies (e.g. in case of war). Strategic adversaries might react by raising their own military expenditures. All this means that the global effects of a military spending shock might not be equal to their domestic effects: if one country's rearmament drive leads to the rest of the world also rearming, the global rise in emissions might be an order of magnitude larger than the domestic effect.

In my sample, there is enough data from 1970 on to look at the effect of US military spending shocks on its strategic allies, which I define as those countries that were members of NATO in 1982.\footnote{Eastern European countries became allies after 1989 after having been adversaries in the previous decades - hence, I do not consider them in this exercise.} That is, I look at the effects of US military expenditure shocks on the military expenditure share of Belgium, Canada, Denmark, France, Germany, Greece, Italy, the Netherlands, Norway, Portugal, Spain, Turkey, and the United Kingdom, using the original local projection setup (2), without time fixed effects (which would filter out the common shock). As the US was one of the two most important military powers in this sample, this is a reasonable setup - I do not have data on Russian military expenditures separate from the USSR before 1991.

\begin{figure}[h!]
    \centering
    \includegraphics[width=0.5\textwidth]{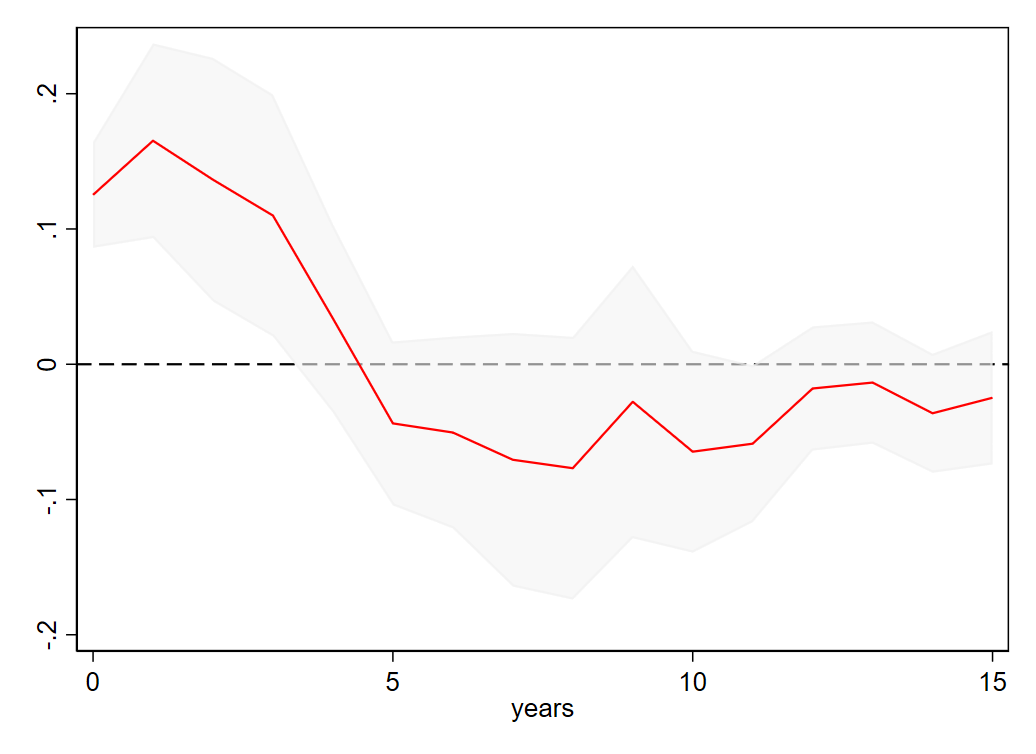}
    \caption{Spillover effect of US shocks}
    \label{fig:emp15}
    \caption*{Note: the figure shows the IRFs of the military spending share in 1982 NATO countries in response to a US military spending shock corresponding to a 1 percentage point rise in the military spending share. The outcome is in terms of percentage changes. The dark grey area marks the 68\% confidence interval.}
\end{figure}

The results can be seen on figure \ref{fig:emp15}. In response to a 1 p.p. rise in the US military spending share, NATO allies raise their military spending share by 0.1-0.2 percentage points in the first 4 years after the shock, following which it returns to what it would have been absent the shock. This suggests that the US's allies generally "showed solidarity" with the US following spending shocks, in the sample. This would suggest that the effects of a military spending share shock in the US would have important positive spillover effects in its NATO allies, meaning that my results might underestimate the global effects.

I do not present results for the strategic adversaries of the US. The reason for this is that there are only two countries in the sample which can be reasonably categorized as adversaries of the US throughout the sample: China and Russia, and I do not have data on Russia's military expenditures before 1990. It would be important for future work to estimate empirically spillover effects on strategic adversaries.

\subsection{Discussion of empirical results}

The empirical results robustly indicate that both emissions and emission intensity increases following a military spending shock equivalent to doubling the military spending share. In the main (NATO) sample, total emissions rise by 1.94\%, while emission intensity rises by 0.9\%. The data suggests that the rise in emissions is due to both a rise in emission intensity and a rise in output following a military spending shock. In the high-emission-intensity countries of the extended sample emissions and intensity also rise, the former by 0.9\%, the latter by more than 1\%. However, low-intensity countries experience a drop in total emissions. I interpret this as evidence that emissions rise by 0.9-2\% in response to a doubling of the military expenditure share, while emission intensity rises by 1\%, with the qualification that this effect might be lower in low energy and emission intensity economies.

The results also suggest that a military spending shock reduces green patenting by 10-25\%, either due to crowding out innovation in this field or as a result of a general negative effect on patenting.

Note that there are other potential margins of heterogeneity that I do not address in my method. As Ohanian (1997) pointed out, financing wars with debt can lead to a more pronounced rise in output than financing it with taxes. An upshot of this is that emissions could also rise more in the case of debt-financing.\footnote{Unless financing methods affect emission intensity differentially.} It is unclear how the defence shocks in the data were financed - this might impact the overall results.

An important difference might arise between temporary and permanent military spending shocks. My empirical strategy was aimed at uncovering the effect of temporary shocks. However, as Barro (1981) pointed out, permanent government spending shocks might affect the economy differently compared to temporary shocks - in particular, their effect on output might be lower.\footnote{A permanent military expenditure shock could materialize if, for example, the international security situation is perceived to worsen in the long run.} In consequence, this would mean that the effect on total emissions in any given year of a permanent shock would be lower.

To cross-check the empirical results and to better understand the mechanisms through which a military spending shock affects emissions and the emission intensity, as well as the consequences of such a shock on the green transition, I use a dynamic production network model.

\section{Model}

This section describes the theoretical model. I use it to provide additional evidence on the effect of a military buildup on greenhouse gas emissions, to understand through what other channels it impacts the green transition, as well as the mechanisms through which these effects work.

\subsection{Setup}

I use a dynamic production network model with an investment network, based on vom Lehn and Winberry (2021). Production networks allow for more precise modelling of inter-industry substitution possibilities and propagation, which is crucial for measuring the change in emissions in response to a military demand shock. The granularity of the industry structure also permits analysing the sectoral effects of a rise in military spending, as well as decomposing the total effect into different components (I discuss these below). In addition, investment dynamics (together with an investment network) are essential for capturing the mechanism of capacity expansion in defence industries, in response to a demand shock.

I do not discuss innovation in the context of my model. The primary reason is that that would make it significantly more complicated, as it would require modelling the market power of the industries and taking a stance on the relevant innovation process, which is beyond this paper's scope.

The model is fully deterministic. There are three types of entities in the model: a representative household, n industries, and the government. The representative household's problem is:
\begin{gather*}
    \max \sum_{t=0}^{\infty}\beta^{t}\Bigg[\log C_{t} - \frac{L_{t}^{1 + 1/\xi}}{1 + 1/\xi}\Bigg]\\
    \text{s.t. } C_{t} = \prod_{i=1}^{n}f_{it}^{\beta_{i}}\\
    \sum_{i=1}^{n}p_{it}f_{it} + \sum_{i}p_{it}^{I}I_{it} = w_{t}L_{t} + \sum_{i}r_{it}K_{it} - T_{t}\\
    K_{it+1} = (1-\delta_{i})K_{it} + I_{it}\\
    I_{it} = \prod_{j}i_{jit}^{\chi_{ji}}\\
    \sum_{i = 1}^{n} \beta_{i} = 1; \text{ } \sum_{j = 1}^{n} \chi_{ji} = 1, \text{ } \forall i
\end{gather*}
Where $C_{t}$ is aggregate consumption in period t, $L_{t}$ is aggregate labour supply, and $f_{it}$ is the household consumption of good variety i in period t. For simplicity, I assume log consumption utility, and an additively separable labour disutility function, where the Frisch elasticity of substitution is $\xi$. I assume a Cobb-Douglas consumption aggregator for simplicity. 

The household accumulates capital specific to industry i ($K_{it}$) by buying investment goods specific to industry i, $I_{it}$ (also a Cobb-Douglas aggregate of the product of different industries). Depreciation of industry i capital is equal to $\delta_{i}$. The representative household buys consumption and investment goods using labour and capital income ($w_{t}$ is the wage, and $r_{it}$ is the return on industry i's capital in period t). $T_{t}$ denotes a lump-sum tax. I set $P_{t}C_{t} = \sum_{i=1}^{n}p_{it}f_{it}$, and I normalize $P_{t} = 1$. All markets - goods, labour, and capital - are perfectly competitive.

The profit maximization problem of industry i is:
\begin{gather*}
    \max p_{it}y_{it} - w_{t}L_{it} - r_{it}K_{it} - \sum_{j} p_{jt}x_{jit}\\
    \text{s.t. } y_{it} = (K_{it}^{1-\alpha_{i}}L_{it}^{\alpha_{i}})^{\theta_{i}}X_{it}^{1-\theta_{i}}\\
    X_{it} = \prod_{j=1}^{n}x_{jit}^{\omega_{ji}}\\
    \sum_{j = 1}^{n} \omega_{ji} = 1, \text{ } \forall i
\end{gather*}
That is, industries produce with a Cobb-Douglas technology, using labour, capital, and intermediate inputs. The aggregate basket of intermediate goods is $X_{it}$, and $x_{jit}$ denotes input bought from industry j and used by industry i.

The government budget constraint is:
\begin{gather*}
    \sum_{i}p_{it}G_{it} = T_{t}
\end{gather*}
That is, in the baseline model, the government simply taxes lump-sum the representative household, and buys goods from different industries ($G_{it}$ denotes government consumption). 

The goods and labour market clearing conditions are:
\begin{gather*}
    y_{it} = f_{it} + G_{it} + \sum_{j} i_{ijt} + \sum_{j} x_{ijt}, \text{ } \forall i\\
    \sum L_{it} = L_{t} 
\end{gather*}
Now, one must define a military buildup in the context of the model. In general, military purchases are quite diverse, ranging from weapons and fuel to food and medicine, as well as salaries and pensions. While all of these are potentially important, I focus on weapon procurements and energy/fuel use, as these are the most likely to raise total GHG emissions and the emission intensity of an economy.\footnote{Another potential candidate would be spending on construction (new bases, storage facilities, etc.) - but it's highly uncertain how likely that is to occupy a significant part of extra military spending during a buildup.}

Hence, the aim is to understand how a positive shock - permanent or temporary - to $G_{it}$, where i represents weapon and energy/fuel industries, impacts total emissions.\footnote{In my empirical analysis, I focused on temporary shocks to be able to extract causal effects. Here, the model permits me to analyse the effects of a permanently higher level of military spending. See below.} \footnote{The shocks are calibrated to match actual military budget compositions - more on this below.} For simplicity, I assume that all extra government spending is funded through lump-sum taxing the household (i.e. the government does not change its purchases of non-shocked goods).

I transform the system into exact-hat form to simplify the analysis. That is, I replace all variables $z$ with $\hat{z}_{t} = \frac{z_{t}}{\Bar{z}}$, where $\Bar{z}$ is the initial steady-state of the model. 

The system of first-order conditions, in exact-hat algebra, is:
\begin{gather}
    \hat{f}_{it} = \hat{p}_{it}^{-1}\hat{C}_{t}\\
    \hat{P}_{t} = \prod_{i=1}^{n}\hat{p}_{it}^{\beta_{i}} = 1\\
    \hat{L}_{t} = \bigg(\frac{\hat{w}_{t}}{\hat{C}_{t}}\bigg)^{\xi}\\
    \hat{i}_{ijt} = \hat{p}_{i}^{-1}\hat{p}_{jt}^{I}\hat{I}_{jt} \\
    \hat{p}^{I}_{it} = \prod_{j}\hat{p}_{jt}^{\chi_{ji}}\\
    \hat{p}_{it}^{I}\hat{C}_{t}^{-1} = \beta \hat{C}_{t+1}^{-1} \Bigg(\frac{1-\beta(1-\delta_{i})}{\beta}\hat{r}_{it+1} + (1-\delta_{i})\hat{p}_{it+1}^{I}\Bigg)\\
    \hat{K}_{it+1} = (1-\delta_{i})\hat{K}_{it} + \delta_{i}\hat{I}_{it}\\
    \hat{y}_{it} = (\hat{VA}_{it})^{\theta_{i}}\hat{X}_{it}^{1-\theta_{i}}\\
    \widehat{VA}_{it} = \hat{K}_{it}^{1-\alpha_{i}}\hat{L}_{it}^{\alpha_{i}}\\
    \widehat{VA}_{it} = \hat{P}_{VAit}^{-1}\hat{p}_{it}\hat{y}_{it}\\
    \hat{P}_{VAit} = \hat{r}_{it}^{1-\alpha_{i}}\hat{w}_{t}^{\alpha_{i}}\\
    \widehat{K}_{it} = \hat{r}_{it}^{-1}\hat{P}_{VAit}\hat{VA}_{it}\\
    \widehat{L}_{it} = \hat{w}_{t}^{-1}\hat{P}_{VAit}\hat{VA}_{it}\\
    \hat{X}_{it} = \hat{P}_{x_{i}}^{-1}\hat{p}_{it}\hat{y}_{it}\\
    \hat{P}_{x_{it}} = \prod_{j=1}^{n}\hat{p}_{jt}^{\omega_{ji}}\\
    \hat{x}_{jit} = \hat{p}_{jt}^{-1}\hat{P}_{x_{it}}\hat{X}_{it}\\
    \hat{y}_{it} = \sum_{j}\Delta_{ij}\hat{x}_{ijt} + \sum_{j}\iota_{ij}\hat{i}_{ijt} + \phi_{i}\hat{f}_{it} + \gamma_{i}\hat{G}_{it}\\
    \sum_{i}\lambda_{i}\hat{L}_{it} = \hat{L}_{t}
\end{gather}
Where I denote as $VA_{it}$ the value-added bundle of firm i in period t, and as $P_{VAit}$ its price. $P_{x_{it}}$ is the price of the intermediate input bundle of firm i in period t. We also have that (recalling $\Bar{z}$ denotes the value of $z$ in the initial steady-state):
\begin{gather}
    \Delta_{ij} = \frac{\Bar{p}_{i}\Bar{x}_{ij}}{\Bar{p}_{i}\Bar{y}_{i}}\\
    \iota_{ij} = \frac{\Bar{p}_{i}\Bar{i}_{ij}}{\Bar{p}_{i}\Bar{y}_{i}}\\
    \phi_{i} = \frac{\Bar{p}_{i}\Bar{f}_{i}}{\Bar{p}_{i}\Bar{y}_{i}}\\
    \gamma_{i} = \frac{\Bar{p}_{i}\Bar{G}_{i}}{\Bar{p}_{i}\Bar{y}_{i}}\\
    \lambda_{i} = \frac{\Bar{w}\Bar{L}_{i}}{\Bar{w}\Bar{L}}
\end{gather}
The detailed derivations are found in the appendix. I assume that the emission intensity of different industries does not change (that is, the amount of emissions per unit of real production). Then, I use the following formula:
\begin{gather*}
    \hat{E} = \sum_{i=1}^{n}\hat{y}_{i}\epsilon_{i} + \hat{C}\epsilon_{HH} + \hat{G}_{fuel}\epsilon_{Govt}
\end{gather*}
Where $E$ denotes total emissions, $\epsilon_{i}$ denotes the share of industry i's scope 1 emissions\footnote{"Scope 1 emissions are direct greenhouse (GHG) emissions that occur from sources that are controlled or owned by an organization (e.g., emissions associated with fuel combustion in boilers, furnaces, vehicles)" (US EPA)} out of the total emissions of the economy, $\epsilon_{HH}$ denotes the share of household scope 1 emissions out of total emissions, and $\epsilon_{Govt}$ is the share of the government's scope 1 emissions out of the total.

\subsection{Calibration}

I calibrate the model using data from the United States. First, I use the BEA industry-to-industry use table for 2017 to calibrate $\Delta_{ij}$, $\phi_{i}$, $\gamma_{i}$, and $\lambda_{i}$, following equations (20)-(24).\footnote{I contract the 71 industry dataset into 41 industries to match the Vom Lehn and Winberry (2021) data (more on this below).} \footnote{Note that the BEA does not provide industry-to-industry use tables. Hence, I invert the industry-to-industry total requirements tables using the method described by Acemoglu et al. (2012).} I also use it to calibrate $\alpha_{i}$, $\theta_{i}$, and $\omega_{ij}$ according to:
\begin{gather*}
    \alpha_{i} = \frac{\Bar{w}_{t}\Bar{L}_{it}}{\Bar{r}_{it}\Bar{K}_{it} + \Bar{w}_{t}\Bar{L}_{it}}\\
    \theta_{i} = \frac{\Bar{r}_{it}\Bar{K}_{it} + \Bar{w}_{t}\Bar{L}_{it}}{\Bar{p}_{i}\Bar{y}_{i}}\\
    \omega_{ij} = \frac{\Bar{p}_{i}\Bar{x}_{ij}}{\Bar{P}_{X_{i}}\Bar{X}_{i}}
\end{gather*}
Note that I assume that the economy was in its steady state in 2017.

I also use the 41 sector partition investment network data provided by vom Lehn and Winberry (2021) to calibrate $\chi_{ij}$ and $\iota_{ij}$. These are:
\begin{gather*}
    \chi_{ij} = \frac{\Bar{p}_{i}\Bar{i}_{ij}}{\Bar{p}^{I}_{j}\Bar{I}_{j}}\\
    \iota_{ij} = \frac{\Bar{p}_{i}\Bar{i}_{ij}}{\Bar{p}_{i}\Bar{y}_{i}}
\end{gather*}
Following vom Lehn and Winberry (2021), I average the investment network parameters ($\chi$) over time (that is, I calculate them for all years between 1948 and 2018, and take their average). I calibrate $\iota$ using 2017 data.

Following vom Lehn and Winberry (2021), I add maintenance investment into the investment network. That is, I assume that each industry procures 12.5\% of its total investment from itself - this can be interpreted as "maintenance investment". The reason for this adjustment is that the investment input-output matrix is singular\footnote{Because investment sales are dominated by 4 main hubs, as described by vom Lehn and Winberry (2021).}, and adding diagonal terms ensures its invertibility. Note that this adjustment is not necessary to calculate the steady-state results - the results are virtually the same without the adjustment in the steady-state calculations. However, it is necessary for calculating the dynamic effects of a temporary shock.

For the depreciation rates $\delta_{i}$, I use the fixed assets depreciation data from vom Lehn and Winberry (2021) - for each industry, I use the average of the depreciation rates between 1947-2018. For the subjective discount rate, $\beta$, I set $\beta = 0.98$. For the Frisch elasticity of substitution, $\xi$, I set $\xi = 0.4$, following Reichling and Whalen (2017).

I use emissions data from the Trucost database to calculate the change in industry emissions predicted by the model. I also use data from the US Department of Energy on scope 1 emissions of the Department of Defense to calculate the change in emissions due to military fuel use.\footnote{\url{https://ctsedwweb.ee.doe.gov/Annual/Report/ComprehensiveGreenhouseGasGHGInventoriesByAgencyAndFiscalYear.aspx}} I use data from the Environmental Protection Agency for data on scope 1 household emissions.\footnote{\url{https://www.epa.gov/sites/default/files/2019-04/documents/us-ghg-inventory-2019-main-text.pdf}, "Inventory of US Greenhouse Gas Emissions and Sinks, 1990-2017" (EPA). I only use emissions from passenger cars and the residential sector as household scope 1 emissions.}

It is important to discuss the calibration of the military spending shocks themselves. As discussed before, I will be focusing on shocks to weapon and energy/fuel industries. The weapon industries are NAICS 2007 industry 332, "Fabricated metal products", which includes "Ammunition, arms, and ordnance manufacturing", and industries 3364-3369, "Other transportation equipment", which includes armored vehicle, tank, missile, and aircraft manufacturing. The energy industry is industry 22, "Utilities", as well as industry 324, "Petroleum and coal products".

I want to define the change in emissions as a function of the size of the military spending shock - that is, by how much does the military spending share of GDP change (as a multiple of the previous level).  Now, the procurement multiplier - that is, by how much does government spending on weapon procurement change - is:
\begin{gather*}
    PM = \frac{S_{P} + s_{P}E}{S_{P}}
\end{gather*}
Where $S_{P}$ is the initial GDP share of government spending on weapon procurement, $s_{P}$ is the share of additional total military expenditures spent on procurement, and E is the rise in the total military expenditure share, in percentage points. Similarly, the energy/fuel multiplier (defined analogously) is:
\begin{gather*}
    EM = \frac{S_{E} + s_{E}E}{S_{E}}
\end{gather*}
Where $S_{E}$ is the initial GDP share of government spending on energy/fuel procurement, and $s_{E}$ is the share of additional total military expenditures spent on energy and fuel. Note that both multipliers are a function of M, the total change in the military spending share, and all the parameters can be calculated from the data, except $s_{P}$ and $s_{E}$. I assume that the rest of the extra spending - i.e. a share equal to $1-s_{P}-s_{E}$ - is spent on military personnel.\footnote{Since, as noted above, I assume that all extra spending is funded through lump-sum taxes, a share $1-s_{P}-s_{E}$ is simply transferred back to the household, lump-sum.} \footnote{In fact, the bulk of the non-procurement, non-energy budget of the DoD in 2017 was spent on "Operations and maintenance", which includes food, clothing, medicine, training expenses, etc. I ignore these in my calibration.} Then, for a shock of size M to the total military spending share, I set:
\begin{gather*}
    \hat{G}_{it} = 
    \begin{cases}
        1 + \rho^{t-t_{0}}(PM-1) & i \in \text{Weapon industries}\\
        1 + \rho^{t-t_{0}}(EM-1) & i \in \text{Energy/fuel industries}\\
        1 & \text{ otherwise}
    \end{cases}
\end{gather*}
Where $t_{0}$ is the date when the shock hits. For a permanent shock, I set $\rho = 1$. For a temporary shock, to roughly match the average persistence of a military spending shock on a 10-year horizon estimated in the empirical analysis (see figure 3), I set $\rho = 0.86$.

Since there is uncertainty regarding the share of potential future defence spending shocks going to equipment and energy, I consider three different calibrations - or scenarios - for $s_{P}$ and $s_{E}$. The baseline scenario has $s_{P} = 0.3$ (30\%) and $s_{E} = 0.05$ (5\%), which is higher than the 2017 share of spending on procurement and energy (around 20\%, and around 2\%, respectively (DoD (2016) and Crawford (2019)). This envisions a scenario where the government focuses on improving the striking power of the military by expanding its material base (number of weapons, ammunition, etc.) and by picking up the pace of military activity (e.g. through more frequent training or naval/air patrols) relative to the peacetime norm. Another reason why a relatively larger share of extra spending would be spent on fuel and weapons is that a war breaks out, and tanks, planes, etc. need to be used more intensively. This leads to a relatively higher rate of fuel use and a higher replacement rate of weapons.\footnote{Cottarelli and Virgadamo (2024) argue that EU countries will have to significantly increase the share of military spending on equipment, infrastructure, and operations to have well-equipped and prepared armed forces. In fact, they show that in the EU, 46.2\% of the increase in defense expenditure since 2015 is due to expenditure on equipment and infrastructure.}

The "personnel-oriented policy" calibration has $s_{P} = 0.2$ and $s_{E} = 0.02$ - that is, the actual share of the 2017 budget spent on procurement and energy/fuel. This scenario assumes that it is going to be very hard for militaries to find enough manpower, and will need to spend a relatively large part of their extra budget on paying its extra personnel. The "material-oriented policy" calibration assumes $s_{P} = 0.4$ and $s_{E} = 0.1$ - this scenario envisions a military that does not have difficulties with personnel and can focus its budget on military materials (fuel and weapons).

\subsection{Results}

\subsubsection{Permanent shock}

First, I consider an exercise where the military spending share is permanently higher. In this context, I simulate the model and calculate the new steady state following the shock. The results can be seen below.

\begin{table}[h]
    \centering
    \begin{tabular}{lccc} \hline
    Calibration & Total emissions & Emission intensity & Real GDP \\
    \hline
    Material-oriented & +1.81\% & +1.5\% & +0.31\% \\
    Baseline & +0.9\% & +0.68\% & +0.22\% \\
    Personnel-oriented & +0.36\% & +0.22\% & +0.14\% \\ \hline
    \end{tabular}
    \caption*{Table 2: Change in total emissions, emission intensity, and real GDP in response to a permanent, 1 p.p. rise in the military spending share under the different shock calibrations.}
    \label{table2}
\end{table}

Table 2 shows the response of GHG emissions and emission intensity in the new steady-state, following a permanent 1 p.p. shock to the military spending share. Total emissions rise by between 0.36\% and 1.81\% (depending on the shock composition), while emission intensity rises by 0.22-1.5\%. Four important points should be made here. First, the rise in emission intensity makes up 75-80\% of the total rise in emissions, which means that, in the model, the total effect is not simply due to a positive spending multiplier. Second, real GDP marginally increases under all calibrations, the fundamental reason being that labour supply increases following a rise in the lump-sum tax, due to the income effect induced by the higher tax.\footnote{Changes in real GDP are calculated using a Laspeyres quantity index. Vom Lehn and Winberry (2021) use a Törnqvist index in their simulations.}

Third, there is a significant overlap between the baseline empirical estimate of an effect of 0.9-2\% and the model results, although the interval is shifted somewhat downwards. Fourth, the results are sensitive to the shock calibration - there is a difference of 1.45\% between the effect in a personnel-oriented and a material-oriented scenario.

\begin{figure}[h!]
    \centering
    \includegraphics[width=0.6\textwidth]{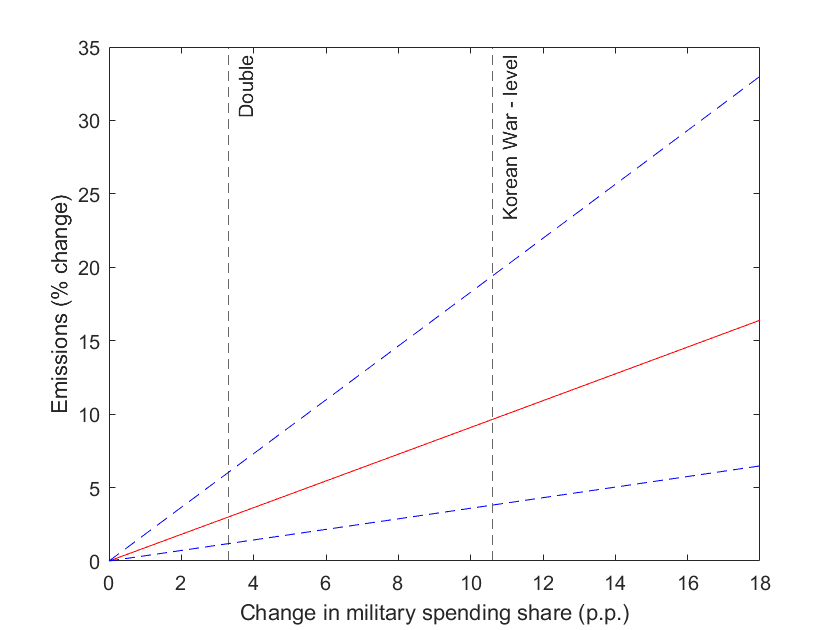}
    \caption{Effect of a permanent military spending shock on total emissions}
    \label{fig:SS}
    \caption*{Note: the figure shows how much total emissions change in the new steady state with the baseline calibration, relative to the old one, in response to a military spending shock (in percentage terms). The size of the spending shock, in percentage points, is shown on the x-axis. The blue dashed lines show the results in the previously defined "personnel-oriented" calibration (lower line), and the "material-oriented" calibration (upper line). Two scenarios are marked with vertical dashed lines: the first is the point at which the military spending share is doubled, and the second is where it is increased to its Korean War peak level.}
\end{figure}

To underscore this uncertainty, figure \ref{fig:SS} shows the percentage change in total emissions, as a function of the size of the military spending share shock (in percentage points). The red line shows the baseline calibration, while the two blue lines show the "personnel-oriented" and the "material-oriented" calibrations. I also show two relevant scenarios with vertical dashed lines: the first shows the point at which the US government doubles its military spending share (from a baseline of 3.3\% in 2017), while the second shows a scenario where the military spending share rises to its Korean War peak (13.9\% of GDP).

In response to a doubling of the military spending share, total emissions rise by approximately 3\% in the new steady state under the baseline calibration. However, depending on the calibration of the shock, this can be both higher or lower. Under the "personnel-oriented" calibration, total emissions rise by only 1.18\%, while under the "material-oriented" policy they rise by 6\%. Also, note that under the extreme case of the military expenditure share rising to its Korean War peak, the baseline calibration predicts an almost 10\% rise in total emissions, with an even wider "confidence interval" - under the material-oriented calibration, emissions rise by 20\%.

From a military point of view (see e.g. Cottarelli and Virgadamo (2024)), the higher the share of expenditures spent on equipment and training, the more efficient and powerful the armed forces are. Hence, the material-oriented scenario better describes a scenario in which, for example, European countries can efficiently invest and improve their militaries. However, in the case of political inefficiencies in constructing and executing defence budgets, an economy can be closer to the personnel-oriented scenario, with small effects on GHG emissions.

\begin{table}[h]
    \centering
    \begin{tabular}{lccc} \hline
    Shocks & Emission intensity & Real GDP & Total effect \\
    \hline
    Weapon procurement & -0.2\% & +0.18\% & -0.02\% \\
    Energy/fuel use & +0.88\% & +0.04\% & +0.92\% \\ \hline
    \end{tabular}
    \caption*{Table 3: Change in total emissions, emission intensity, and real GDP due to the weapon procurement and the energy/fuel use shocks, separately. A military expenditure shock of 1 p.p. under the baseline shock calibration is assumed. The results in either row are calculated by switching off the other shock.}
    \label{table4}
\end{table}

Table 3 decomposes the effect of a 1 p.p. military expenditure share shock into the effect of the two components, the weapon procurement shock and the energy/fuel use shock. I do this by shutting down each shock in turn and looking at the effect of the remaining shock on emission intensity and output.

The results show that the primary reason behind the rise in total emissions is the energy/fuel demand shock. The weapon procurement shock decreases total emissions by a small amount - the explanation for this is that, in this model, the representative household reduces its demand for energy and fuel, which more than counteracts the effect of government demand for energy-intensive goods. Real GDP increases for both shocks, which is simply the outcome of the income effect - note that the effect of the weapon procurement shock on real GDP is higher because it makes up 30\% of the total expenditure shock under the baseline calibration, while the energy demand shock is just 5\% of the total.

\begin{table}[h]
    \centering
    \begin{tabular}{lccccc} \hline
    & Primary & Rubber & Oil and gas & Accommodation and & Agriculture \\
    & metals  & and plastic & extraction & food service & \\ \hline
    Doubling & +5.55\% & +0.73\% & +2.52\% & -0.79\% & -0.72\% \\
    Korean War level & +17.76\% & +2.36\% & +8.07\% & -2.51\% & -2.29\% \\ \hline
    \end{tabular}
    \caption*{Table 4: Change in industry-level output in response to doubling or bringing the military expenditure share to its Korean War peak level.}
    \label{table5}
\end{table}

Note that - as table 4 shows - one of the effects of the military spending shock is that, in the new steady state, energy-intensive industries, like primary metal producers, rubber and plastic firms (which all produce materials essential for weapon production), as well as the oil industry experience a significant boost. Services and agriculture lose the most. This could lead to an increase in the lobbying and political power of these energy-intensive and fossil fuel industries, which could make the green transition harder to accomplish (see, e.g., Sullivan et al. (2022) or Pacca et al. (2021)).

\subsubsection{Temporary shock}

Now let us consider a temporary shock to military spending. In case international tensions subside in the medium-run, the military spending share may also slowly return to its long-run trend. This section considers the effects of such a temporary shock. As mentioned previously, the persistence of the shock is set to $\rho = 0.86$, to roughly match the persistence of the empirically identified shock. I solve for the IRFs of the shock by log-linearizing the model.

\begin{figure}
\centering
\includegraphics[width=0.6\linewidth]{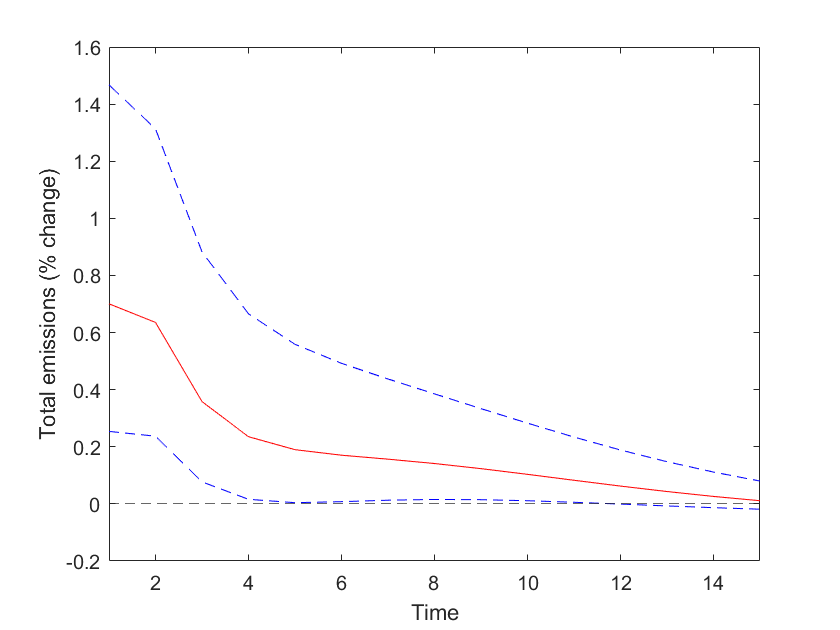}
\caption{IRF of a 1 p.p. temporary military spending share shock's effect on total emissions. The red line shows the baseline calibration, and the upper and lower blue dashed lines show the "material-oriented" and "personnel-oriented" calibrations, respectively.}
\label{fig:tempshock1}
\end{figure}

Figure \ref{fig:tempshock1} shows the effect of the temporary shock of 1 percentage point on total emissions. The contemporary effect in the baseline shock calibration is approximately 0.7\% - somewhat less than the long-run effect of a military spending shock. Following that, emissions slowly return to their long-run trend. In case new spending is primarily material-oriented, this effect is much higher (almost 1.5\% contemporaneously) for longer, while if it's personnel-oriented, it is lower, and returns to virtually zero after 4 years.

\begin{figure}
\centering
\includegraphics[width=0.6\linewidth]{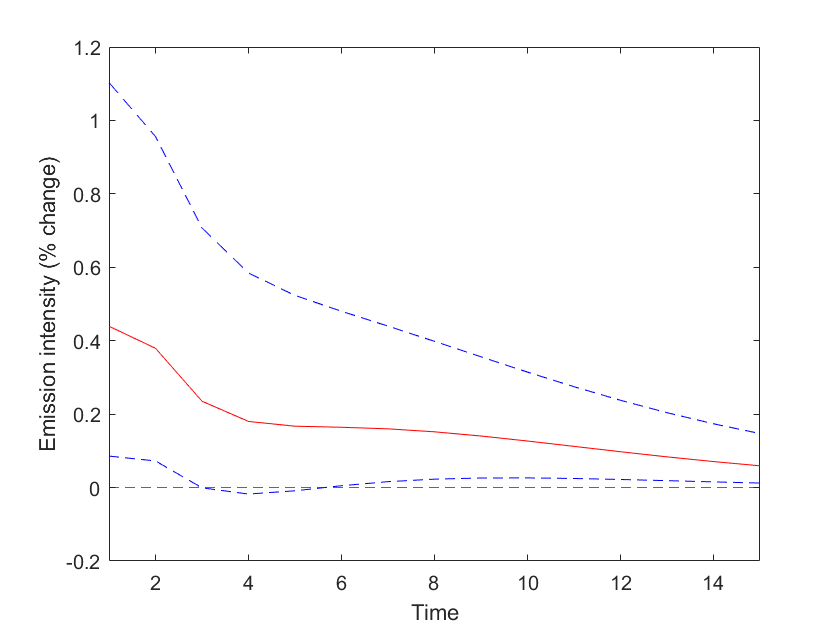}
\caption{IRF of a temporary military spending shock's effect on emission intensity. The red line shows the baseline calibration, and the upper and lower blue dashed lines show "material-oriented" and "personnel-oriented" calibrations, respectively.}
\label{fig:tempshock2}
\end{figure}

Figure \ref{fig:tempshock2} shows that this effect is caused, primarily, by a rise in emission intensity: emission intensity rises by more than 0.4\% contemporaneously for the baseline calibration, about 57\% of the total effect.

\begin{figure}
\begin{subfigure}{.5\textwidth}
  \centering
  \includegraphics[width=1\linewidth]{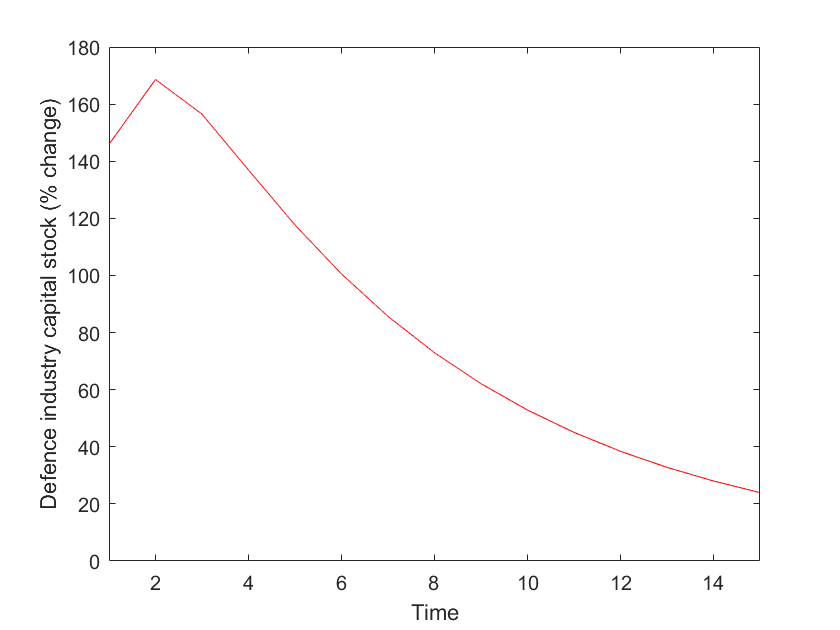}
  \caption{Defence industry - capital}
  \label{fig:sfig13}
\end{subfigure}%
\begin{subfigure}{.5\textwidth}
  \centering
  \includegraphics[width=1\linewidth]{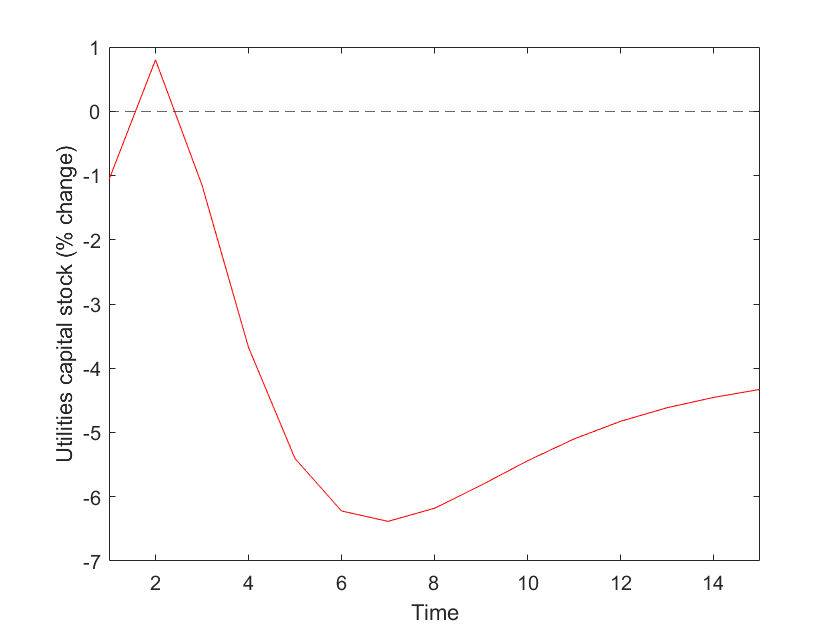}
  \caption{Utilities - capital}
  \label{fig:sfig23}
\end{subfigure}
\caption{IRF of the effect a temporary military spending shock, which brings the spending share to its Korean War peak level, on industry-level capital stocks, under the baseline calibration. Figure a) shows the capital stock in the "Other transport equipment" industry, which contains a large part of the defence industry, while figure b) shows the utilities industry.}
\label{fig:tempshock3}
\end{figure}

The mechanism is broadly similar to the one identified in the previous section, with a permanent shock. It is important, however, to note that the contemporary effect is also driven by a large investment shock, which mostly disappears as capital is reallocated to accommodate the new, defence-oriented economic structure.\footnote{This temporary investment boom is partly responsible for the contemporary rise in the measured overall production, mentioned previously.} In particular, figure \ref{fig:tempshock3} shows the evolution of the capital stock in two different industries: the "Other transport equipment" industry, which makes up a large fraction of the defence industry, and the power sector, in response to a shock that brings the military spending share to its Korean War peak level. Investment booms in the defence industries, with the capital stock peaking at more than 160\% above its trend level 2 years after the shock.\footnote{The IRFs show the evolution of the capital stock used in the next period - that is, the value of the IRF in period t reflects the movement in $K_{t+1}$.} Meanwhile, the capital stock in the utilities sector stagnates in the first two periods, after which investment slumps, and the capital stock falls by 6\% below trend, 7 years after the shock. In general, investments in the defence industries crowd out investments in other sectors of the economy.\footnote{Crowd-out of investment in utilities happens even though the military spending shock has a positive effect on it. As lump-sum taxes are used to fund new expenditures, the household reduces its spending on energy significantly. In addition, some other industries which are negatively affected also reduce their energy demand.}

The reason why it is particularly important to emphasize the slump in investment in the utilities sector is that new investments in this industry usually go to renewable projects (see IEA (2023): investment in renewables was more than half of all investments in the power sector globally, while investment in fossil fuels was less than 10\%) - hence, a crowd-out of investment from this sector could retard the green transition by slowing down investment in green energy sources. Note, though, that investment in the power sector increases with a permanent shock in the new steady state. This suggests that whether investment in renewables falls or not might depend on the persistence of the rise in military spending. However, in case investment in the power sector increases in a new steady state, it is unclear whether the extra investment will flow to renewables, electrical grids, or storage technologies, or whether it will flow to dirty plants.

Note that the empirical IRF features a non-contemporaneous peak in total emissions and emission intensity. The reason why this is not replicated in the model results is that, in the model, new military spending and production happen simultaneously, while in reality there could be a lag in production, as discussed above. It is also important to mention that governments could finance extra military spending with deficits, which could lead to a higher level of overall production and, thus, to a larger volume of total emissions (reflecting the result by Ohanian (1997) - although note that he considered distortionary taxation, not lump-sum taxes).

To sum up: the model predicts that both total emissions and the emission intensity rise in response to a military spending shock. A permanent shock to the military spending share of 1 percentage point raises total emissions by between 0.36\% and 1.81\%, depending on the composition of the military spending shock. It also raises emission intensity by between 0.22\% and 1.5\%. A temporary shock similarly raises total emissions and emission intensity persistently.

Energy-intensive industries, as well as fossil-fuel industries, experience a significant boost in size and activity, while investment and production in other sectors are crowded out. This can lead to a fall in the investment in renewables, which could slow the green transition.

\section{The Green Peace Dividend and policy implications}

This section aims to summarize the overall results regarding the impact of a military buildup on emissions and the green transition. As military buildups have several different effects that are potentially relevant, many of them hard to quantify, I aim to describe qualitatively the climate benefits of peace - the Green Peace Dividend (henceforth GPD). Some of these, however, yield themselves to quantification, in which case I provide numbers that can serve as useful reference points.

First of all, it is most straightforward to quantify the effect of military buildups on GHG emissions. This is what the empirical analysis and most of the model analysis was focused on. The empirical analysis indicates that a one percentage point rise in the military spending share leads to a rise in total emissions of between 0.9 and 2\%, depending on the method and sample used. The model indicates, under the baseline calibration for the US in 2017, that a permanent one percentage point rise in the military spending share raises GHG emissions by between 0.36\% and 1.81\%, depending on the composition of military spending.

The results suggest that the rise in GHG emissions caused by a militarization episode is significant. It is also reassuring that the empirical and theoretical results are similar. However, it is worth keeping in mind that the results depend on the specific characteristics of the country of interest - for example, the empirical analysis shows that low-emission-intensity countries might not experience a rise in emissions in response to such a shock.

I use two main methods to assess the impacts of the measured effects on GHG: by using the social cost of carbon (SCC), and by looking at the global path of surface temperatures.

First, focusing solely on these direct effects of military buildups on GHG emissions, it is possible to translate the damages due to induced emissions into dollar terms using the social cost of carbon (SCC). For example, if one uses the social cost of carbon (SCC) provided by the EPA (2023), equal to \$190 per ton of CO2 at a near-term discount rate of 2\%, it is possible to consider a scenario in which the US permanently doubled its military spending share in 2017 and to calculate the climate damages resulting from this policy change using the SCC. Total GHG emissions in the US in 2017 were equal to 6.09 Gt of CO2e. The model estimate of a 1.18-6\% rise in emissions in response to such a shock (with a military spending share of 3.3\% in 2017) leads to global climate damages equivalent to approximately 13.6-69.4 billion dollars per year - or approximately 0.07-0.35\% of 2017 US GDP.\footnote{This calculation ignores the transition path to the new steady-state, as well as the fact that emissions change from their 2017 level for reasons independent of defence policy.} More recent - and larger - estimates, such as the SCC calculated by Bilal and K\"anzig (2024), equal to \$1367 per ton of CO2, would yield an interval of 98.2-500 billion dollars per year - or approximately 0.5-2.6\% of 2017 US GDP. Note that these simple calculations ignore the convexity of climate damages.

Second, one can check how a global military buildup of 3.3 percentage points would change the path of temperatures, under different shared socioeconomic pathways (SSP) (Riahi et al. (2017)). Here, I use the central estimate of the effect of a military expenditure shock on emissions from the model, which implies that a permanent shock of 3.3 p.p. raises GHG emissions by 3\%. Using \url{live.magicc.org}, an online software relying on MAGICC7 (Meinsshausen et al. (2020)) (which is an integrated assessment model), I calculate that the difference would be minimal under the SSP1-1.9 scenario (which assumes that GHG emissions are very low, and are reduced to zero by 2050), with temperatures being higher by approximately 0.01 \degree C in 2100 (see figure 19).\footnote{In the simulations, I set CO2 emissions from fossil fuel use and industrial activity to be higher by 3\%, after 2015.} However, under the SSP3 scenario (where GHG emissions double by 2100), temperatures would be higher by 0.05 \degree C by 2100, indicating that the effect of a military buildup becomes worse under worse scenarios. Under the SSP5 scenario, temperatures would be higher by more than 0.06 \degree C.\footnote{For an explanation of the different shared socioeconomic pathways, see: \url{https://www.carbonbrief.org/explainer-how-shared-socioeconomic-pathways-explore-future-climate-change/}} This indicates that under progressively worse scenarios, the effect of a military buildup on temperatures also becomes stronger, as a 3\% rise in a higher baseline amount of emissions leads to a larger change in the level of emissions, leading, in turn, to a larger change in the level of the temperature. Note that temperatures in 2100 rise with the shown scenarios (so they are highest under SSP5, and lowest under SSP1 - 1.9), and temperatures peak before 2100 only under the SSP1 scenarios.

\begin{figure}[h!]
    \centering
    \includegraphics[width=0.6\textwidth]{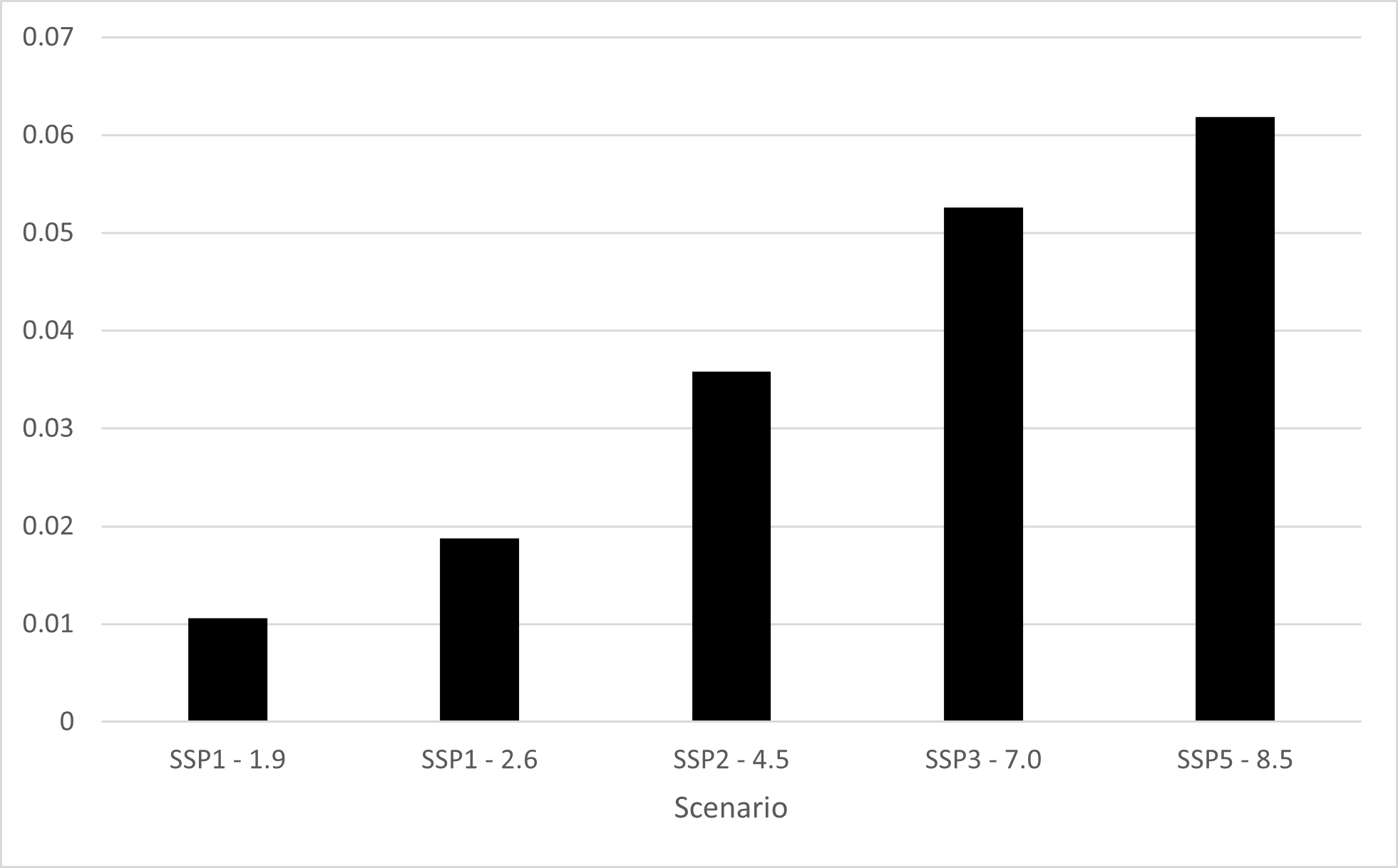}
    \label{fig:SSP}
    \caption{Effects on temperature under different SSP scenarios}
    \caption*{Note: the figure shows by how much higher surface temperatures would be in 2100 given that a permanent global military buildup of 3.3 p.p. happens, in \degree C, under different shared socioeconomic pathways.}
\end{figure}

These damage calculations omit several effects which cannot be properly quantified. First of all, the empirical results show that green patenting falls significantly following a military spending shock. This could, in itself, retard the transition to net zero emissions. However, if fossil fuel and energy-intensive industries do not reduce or even scale up "dirty" or "grey" R\&D expenditures, this could contribute to a widening gap relative to clean technical progress - as Acemoglu et al. (2012) argued, such a gap could be difficult to overcome, and could be costly to close.

Second, the model indicates that investments in energy-intensive industries crowd out investments in other sectors after a military spending shock. In particular, investment in the utilities sector slumps under a temporary military spending shock, which is worrying because this can translate into a slump in renewables investments, as the majority of investments in this sector are directed towards renewables.

Third, both the empirical and the model results show that emission intensity increases significantly in response to a military buildup. The reason for this, as the model analysis showed, is that energy-intensive and fossil fuel industries experience an expansion. As argued previously, this can lead to a rise in their political influence and lobbying power, which could make passing carbon policies and the green transition much more complicated.

Also, note that if there indeed is a causal effect of climate change on the risk of conflict, then a positive feedback loop generated by the impact of military buildups on climate change increases the damages due to both of these phenomena.

While it is hard to quantify the exact impact of all these consequences, it seems clear that they could pose a major obstacle to the green transition. To avoid that in case a military buildup becomes necessary, the government must balance these negative effects with additional policies - I discuss some possibilities in the next section.

I do not discuss wars specifically here. Wars lead to additional emissions - for example, through forest fires caused by the militaries (think of the napalm bombing campaigns in Vietnam), or indirectly, through destroying buildings that need to be rebuilt later. de Klerk et al. (2023) and Neimark et al. (2024) provide detailed analyses of the environmental impacts of the Ukraine war and the Israel-Hamas war, respectively.

To sum up, military buildups impact the carbon budget and cause climate damages directly by increasing GHG emissions. In addition, they might significantly complicate the green transition by crowding out green innovation and investment and by bolstering fossil-fuel and energy-intensive industries that could obstruct the passing and implementation of climate policies.

\subsection*{Policy}

The primary issue that climate policymakers must face when coming up with a response to the negative impacts of militarization is that defence policy decisions are almost completely orthogonal to climate policy. In addition, military buildups, by deterring enemy action, can promote peace - and deciding to wage a war can, in some cases, reduce the chances of larger conflagrations in the future, or avoid disastrous long-run consequences such as military occupation, extractive dictatorships or even genocide. Hence, avoiding militarization is not necessarily in the interest of climate policymakers.

There are, broadly, two possible paths - or a combination of them - that policymakers can follow to avoid the worst consequences of military buildups on the climate. First of all, it is possible to tighten existing carbon taxes or caps to counteract the rise in emissions due to a higher level of military spending. While this might, in theory, be effective, there are several problems with implementing it in practice.

Military buildups, war preparations, and wars usually take place in the context of high social tensions. Preserving social solidarity and minimizing the sacrifices and burdens that the population must bear is essential for the success of such undertakings. Carbon pricing, however, is known to reduce economic activity and to have unequal impacts on income and consumption (see, e.g., K\"anzig (2023)), which imperils both of these objectives. Targeted fiscal policy can, as K\"anzig argues, alleviate these effects, but it is not clear whether a public administration primarily focused on defence policy would have the capabilities to successfully execute such a policy mix. Nevertheless, some additional carbon pricing might still be feasible, dampening the effect of militarization on emissions.

Second, approaching this issue from the side of subsidies and innovation might be more feasible, both from a political point of view, as well as because some useful synergies could arise between climate and defence policy goals. Subsidizing investment in renewables and green R\&D can counteract the crowd-out effect described in the previous sections, although it most likely wouldn't reduce emissions in the short run. In case military spending directly or indirectly leads to a rise in dirty innovation, such a policy could also prevent the gap between dirty and clean technical progress from widening - something which could be costly to overcome in the future, as Acemoglu et al. (2012) argued. 

Investing in renewables can also improve the energy security of a country, especially if it relies on significant energy imports. Shielding household energy prices from foreign energy market volatility could ease the burden that households must bear in a conflict or a rearmament program, especially if international tensions are widespread.

In sum, the carrot - green subsidies - is more likely to be successful in overcoming the negative effects of a military buildup that the stick, carbon pricing. Nevertheless, a combination of the two might be necessary to fully offset these effects and to keep the green transition on track.

\section{Conclusion}

In this paper, I show that military buildups increase greenhouse gas emissions and can significantly complicate the green transition. The empirical analysis suggests that a one percentage point rise in the military spending share increases total emissions by 0.9-2\%, and emission intensity by 1\%. It also suggests that such a shock can reduce green patenting by 10-25\%.

The dynamic production network model, calibrated using US data from 2017, indicates that military buildups increase both total emissions, as well as emission intensity. The magnitude of these effects, however, seems to be lower than the one indicated by the empirical analysis. Permanently doubling the military spending share raises total emissions by between 0.36\% and 1.81\%, while it raises emission intensity by between 0.22\% and 1.5\%. Hence, military buildups can directly reduce our carbon budget. However, the model analysis also indicates that militarization can impede the green transition by crowding out investment in renewables, in the case of a temporary military spending shock. 

My analysis, based on the model results, indicates that doubling the military spending share in the US in 2017 would have led to climate damages equal to between 0.07-2.6\% of GDP per year, depending on the SCC and the shock composition. Importantly, these damage calculations do not take into account potential effects on green innovation or investments.

Policymakers can counteract these negative effects using either carbon pricing or subsidizing green investment and R\&D, or through a combination of these. I argue that subsidies are likely to be more efficient, partly due to political and social constraints, and partly because they can exploit potential synergies between defence and climate policy goals.

\nocite{*}
\bibliography{referencesarxiv.bib} 
\bibliographystyle{apalike} 

\section*{Appendix}

\subsection*{Deriving the equilibrium exact-hat equations}

The household problem's first-order conditions are:
\begin{gather*}
    f_{it} = \beta_{i}p_{it}^{-1}C_{t} \\
    L_{t} = \bigg(\frac{w_{t}}{C_{t}}\bigg)^{\xi} \\
    i_{ijt} = \chi_{ji}p_{i}^{-1}p_{jt}^{I}I_{jt} \\
    p_{it}^{I}C_{t}^{-1} = \beta p_{it+1}^{I}C_{t+1}^{-1} \Bigg(\frac{r_{it+1}}{p_{it+1}^{I}} + (1-\delta_{i})\Bigg)
\end{gather*}
Where the price index of the consumption basket is $P_{t} = \prod_{i=1}^{n}p_{it}^{\beta_{i}}$, which I normalize to one. The sector i investment good's price index is $p^{I}_{it} = \prod_{j}p_{jt}^{\chi_{ji}}$.

As mentioned previously, the goal is to transform all equations into their "exact-hat form" (Dekle, Eaton, and Kortum (2007)). That is, all variables are shown as their value relative to the initial steady-state: for variable x, $\hat{x}_{t} = \frac{x_{t}}{\Bar{x}}$, where $\Bar{x}$ is the initial steady-state value. It is easy, hence, to transform the static equations into this form: one just divides them by their steady-state equivalent. Applying this method to the previously listed household FOCs - except the Euler equation - one gets equations (3) to (7).

For the Euler equation (8), one starts out from its steady-state equivalent:
\begin{gather*}
    \Bar{p}_{i}^{I}\Bar{C}^{-1} = \beta \Bar{p}_{i}^{I}\Bar{C}^{-1} \Bigg(\frac{\Bar{r}_{i}}{\Bar{p}_{i}^{I}} + (1-\delta_{i})\Bigg) \\
    \implies \Bar{p}_{i}^{I} = \frac{\beta}{1-\beta(1-\delta_{i})} \Bar{r}_{i}
\end{gather*}
Hence, dividing the Euler equation in levels by $\Bar{p}^{I}\Bar{C}^{-1}$, we get:
\begin{gather*}
    \hat{p}_{it}^{I}\hat{C}_{t}^{-1} = \beta \hat{C}_{t+1}^{-1} \Bigg(\frac{r_{it+1}}{\Bar{p}_{i}^{I}} + (1-\delta_{i})\hat{p}_{it+1}^{I}\Bigg)
\end{gather*}
And, by substituting in the previous identity, I get equation (7).

We can get equation (9) - the capital accumulation equation - by using its steady-state form:
\begin{gather*}
    \Bar{K}_{i} = (1-\delta_{i})\Bar{K}_{i} + \Bar{I}_{i} \\
    \Bar{I}_{i} = \delta_{i}\Bar{K}_{i}
\end{gather*}
Hence, we have, by dividing through the capital accumulation equation in period t by the steady-state value of investment:
\begin{gather*}
    K_{it+1} = (1-\delta_{i})K_{it} + I_{it} \\
    \hat{K}_{it+1}\delta_{i}^{-1} = (1-\delta_{i})\delta_{i}^{-1}\hat{K}_{it} + \hat{I}_{it}\\
    \hat{K}_{it+1} = (1-\delta_{i})\hat{K}_{it} + \delta_{i}\hat{I}_{it}
\end{gather*}
Now, industry i's first-order conditions are:
\begin{gather*}
    VA_{it} = \theta_{i}P_{VAit}^{-1}p_{it}y_{it} \\
    K_{it} = (1-\alpha_{i})r_{it}^{-1}P_{VAit}VA_{it} \\
    L_{it} = \alpha_{i}w_{t}^{-1}P_{VAit}VA_{it} \\
    X_{it} = (1-\theta_{i})P_{x_{i}}^{-1}p_{it}y_{it}
\end{gather*}
Where the value-added bundle's price index is $P_{VAit} = r_{it}^{1-\alpha_{i}}w_{t}^{\alpha_{i}}$, and the intermediate good bundle's price index is $P_{x_{it}} = \prod_{j=1}^{n}p_{jt}^{\omega_{ji}}$. As these are all static equations, it is, again, straightforward to divide them by their steady-state equivalent and get equations (12) to (18). Similarly, equations (10) and (11) can be reached by dividing the definition of the production function and the value-added bundle by their steady-state equivalent.

The goods market clearing conditions are:
\begin{gather*}
    y_{it} = \sum_{j}x_{ijt} + \sum_{j}i_{ijt} + f_{it} + G_{it}
\end{gather*}
By dividing through with $\Bar{y}_{i}$:
\begin{gather*}
    \hat{y}_{it} = \sum_{j}\frac{\Bar{p}_{i}\Bar{x}_{ij}}{\Bar{p}_{i}\Bar{y}_{i}}\hat{x}_{ijt} + \sum_{j}\frac{\Bar{p}_{i}\Bar{i}_{ij}}{\Bar{p}_{i}\Bar{y}_{i}}\hat{i}_{ijt} + \frac{\Bar{p}_{i}\Bar{f}_{i}}{\Bar{p}_{i}\Bar{y}_{i}}\hat{f}_{it} + \frac{\Bar{p}_{i}\Bar{G}_{i}}{\Bar{p}_{i}\Bar{y}_{i}}\hat{G}_{it}
\end{gather*}
Which gives equation (19). Similarly, the labour market clearing condition is:
\begin{gather*}
    \sum_{i}L_{it} = L_{t}
\end{gather*}
Where, dividing by $\Bar{L}$, we get:
\begin{gather*}
    \sum_{i}\frac{\Bar{w}\Bar{L}_{i}}{\Bar{w}\Bar{L}}\hat{L}_{it} = L_{t}
\end{gather*}
Which gives equation (20).

\end{document}